\documentclass[12pt]{article}
\usepackage{amsmath,amsfonts,amssymb}
\usepackage{hyperref}
\usepackage{graphicx}
\usepackage{color}
\usepackage{epsfig}
\usepackage{psfrag}
\usepackage{tikz}
\usepackage{slashed}
\usepackage{ulem}
\usepackage{comment}

\usepackage{color}

\newcommand \be{\begin{eqnarray}}
\newcommand \ee{\end{eqnarray}}

\makeatletter\@addtoreset{equation}{section}\makeatother

\DeclareMathOperator{\Tr}{Tr}
\DeclareMathOperator{\tr}{tr}
\DeclareMathOperator{\sTr}{sTr}

\newcommand{\ul}[1]{{\underline{#1}}}

\def\bC {\mathbb{C}}

\def\bR {\mathbb{R}}

\def\bZ {\mathbb{Z}}

\newcommand{\bea}{\begin{eqnarray}}
\newcommand{\eea}{\end{eqnarray}}
\newcommand{\beq}{\begin{equation}}
\newcommand{\eeq}{\end{equation}}
\newcommand{\bal}{\begin{equation}\begin{aligned}}
\newcommand{\eal}{\end{aligned} \end{equation}}

\newcommand{\vev}[1]{{\left< {#1} \right>}}

\newcommand{\eqn}[1]{(\ref{#1})}
\newcommand{\tx}[1]{\text{#1}}
\newcommand{\app}[1]{Appendix~\ref{app:#1}}
\newcommand{\sect}[1]{Section~\ref{sec:#1}}

\newcommand{\address}[1]{\vbox{\center\em#1}}
\renewcommand{\title}[1]{\vbox{\center\huge{#1}}\vspace{5mm}}

\newcommand{\cA}{{\mathcal A}}

\newcommand{\cD}{{\mathcal D}}

\newcommand{\cL}{{\mathcal L}}
\newcommand{\cM}{{\mathcal M}}
\newcommand{\cN}{{\mathcal N}}
\newcommand{\cP}{{\mathcal P}}

\newcommand{\cR}{{\mathcal R}}

\newcommand{\cS}{{\mathcal S}}
\newcommand{\cW}{{\mathcal W}}

\newcommand{\mt}[1]{\textrm{\tiny #1}}

\begin{document}
\bibliographystyle{utphys}

\begin{titlepage}
\begin{center}
\phantom{ }
%\vskip5mm

\vspace{5mm}

\title{A profusion of $1/2$ BPS Wilson loops in \\ $\cN=4$ Chern-Simons-matter theories}
\vspace{3mm}

\renewcommand{\thefootnote}{$\alph{footnote}$}

Michael Cooke%
\footnote{\href{mailto:cookepm@tcd.ie}{\tt cookepm@tcd.ie}}, 
Nadav Drukker%
\footnote{\href{mailto:nadav.drukker@gmail.com}{\tt nadav.drukker@gmail.com}} 

%\vskip 2mm
\address{
Department of Mathematics, King's College London \\
The Strand, WC2R 2LS, London, UK}

\vskip 5mm

and Diego Trancanelli%
\footnote{\href{mailto:dtrancan@usp.br}{\tt dtrancan@usp.br}} 
%\vskip 2mm
\address{
Institute of Physics, University of S\~ao Paulo\\
05314-970 S\~ao Paulo, Brazil}

\renewcommand{\thefootnote}{\arabic{footnote}}
\setcounter{footnote}{0}

\end{center}

\vskip5mm

\abstract{
%\normalsize
\noindent
We initiate the study of $1/2$ BPS Wilson loops in $\cN=4$ Chern-Simons-matter theories 
in three dimensions. We consider a circular or linear quiver with Chern-Simons levels 
$k$, $-k$ and $0$, and focus on loops preserving one of the two $SU(2)$ subgroups 
of the $R$-symmetry. 
In the cases with no vanishing Chern-Simons levels, we find a pair of Wilson loops 
for each pair of adjacent nodes on the quiver connected by a hypermultiplet (nodes connected 
by twisted hypermultiplets have Wilson loops preserving another set of supercharges). 
We expect this classical pairwise degeneracy to be lifted by quantum 
corrections. In the case with nodes with vanishing Chern-Simons terms connected by 
twisted hypermultiplets, we find that the usual $1/4$ BPS Wilson loops are automatically 
enlarged to $1/2$ BPS, as happens also in 3-dimensional Yang-Mills theory. When the nodes 
with vanishing Chern-Simons levels
are connected by untwisted hypermultiplets, we do not find any Wilson 
loops coupling to those nodes which are classically invariant. Rather, we find several loops 
whose supersymmetry variation, while non zero, vanishes in any correlation function, so is 
weakly zero. We expect only one linear combination of those Wilson loops to remain 
BPS when quantum corrections are included. We analyze the M-theory duals of those 
Wilson loops and comment on their degeneracy. We also show that these Wilson loops 
are cohomologically equivalent to certain $1/4$ BPS Wilson loops whose expectation 
value can be evaluated by the appropriate localized matrix model. 
}

\end{titlepage}

\setcounter{tocdepth}{2} % Eliminates subsubsections from table of contents
\addtolength{\parskip}{-.4mm}
\tableofcontents
\addtolength{\parskip}{.4mm}

\section{Introduction}
\label{sec:intro}

It is by now a widely substantiated fact that 
BPS Wilson loops provide a powerful probe of supersymmetric gauge theories. 
While Wilson loops in general enable us to study gauge theories and their phases, 
the BPS ones are particularly interesting as they also 
allow to perform exact calculations, via localization \cite{pestun}.
The prototypical example of such operators is the $1/2$ BPS Wilson loop of $\cN=4$ super 
Yang-Mills (SYM) theory in four dimensions, which is defined in terms of a 
gauge connection augmented by a scalar coupling \cite{maldacena,rey,gross}. In the case of the circular 
geometry, this is evaluated by the Gaussian matrix model \cite{erickson,DG}. The same is true for 
a much larger class of theories with $\cN=2$ supersymmetry, where again a scalar 
coupling allows to turn the loop for any gauge group in a quiver gauge theory 
(or a theory of class $\cS$) to be BPS and they provide rich probes of the theory 
\cite{pestun,AGGTV,DGOT,GMN}.

In three dimensions the story is quite similar, if we consider Chern-Simons (CS) theories with $\cN=2$ 
supersymmetry. The $1/2$ BPS Wilson loops there again involve a coupling to a scalar from 
the vector multiplet \cite{yin} and can be evaluated by localization to a 
matrix model \cite{kapustin}. 
The same is true for Yang-Mills theories with $\cN=4$ supersymmetry \cite{assel_gomis}. 
Things are however more complicated when considering Chern-Simons theories with more extended 
supersymmetry. In the ABJ(M) theory \cite{abjm,abj}, 
which has $\cN=6$ supersymmetries, the loops with only scalar couplings 
turn out to be only $1/6$ BPS \cite{Chen:2008bp, plefka, Rey:2008bh}, 
while the $1/2$ BPS ones require a much more complicated structure, involving 
couplings to both gauge groups and the inclusion of fermionic terms \cite{drukker}. Technically, the 
proof of supersymmetry invariance for those 1/2 BPS loops becomes more complicated, since the connection itself is 
not invariant under the supersymmetry variations, but gives a total derivative which needs to be integrated along 
the Wilson loops.

In this paper, we initiate the study of $1/2$ BPS Wilson loops in theories with Chern-Simons 
couplings and $\cN=4$ supersymmetry. The case of CS theories with $\cN=3$ supersymmetry was already 
studied in \cite{chen}, where it was found that the only supersymmetric loops are $1/3$ BPS. The structure 
of $\cN=4$ theories is much more restrictive and we may hope that there could be $1/2$ BPS 
loops. Indeed we shall present an embarrassingly large number of Wilson loops that seem to 
be $1/2$ BPS and will discuss how this degeneracy may be lifted. 

Before outlining the results, we proceed by specifying the theories analyzed in this paper.

%%%%%%%%%%

\subsection{The theories}
\label{sec:thy}

The first $\cN=4$ superconformal Chern-Simons-matter theories
in three dimensions were constructed in \cite{gaiotto}. These were generalized in 
\cite{hoso_4} by the inclusion of ``twisted'' hypermultiplets and further generalized 
in \cite{im_aux}, whose notations we mainly use, see Appendix \ref{app:not} for details. 
The theories we consider are circular or linear quivers with $U(N_I)$ gauge group nodes. 
Adjacent nodes can be connected by bifundamental fields which 
are either hypermultiplets or twisted hypermultiplets. We assume that there are 
$p$ hypermultiplets and $q$ twisted hypermultiplets. A circular quiver has then 
$p+q$ nodes, while a linear quiver has $p+q+1$ nodes.
The supersymmetry enhancement of these extended theories was studied in 
\cite{hoso_6} and, in particular, it was shown that ABJ(M) lies in this class of 
theories.

The vector multiplets associated to the nodes with vanishing CS levels may be 
integrated out to obtain a non-linear theory \cite{koh}. 
We do not adopt this approach and work with the UV description.

We label our nodes by the index $I$, so the vector field is $A_{(I)\mu}$. The gaugino and auxiliary 
scalar (in $\cN=2$ language) are $\lambda_{(I)}$ and $\varphi_{(I)}$, though we will integrate 
them out for nodes with non-vanishing CS levels. The hypermultiplets in the bifundamental representation of 
nodes $I$ and $I+1$ have a scalar $q_{(I)}^{\bar A}$ and fermion $\psi_{(I)\ul B}$. 
They are doublets of the $SU(2)_A$ and $SU(2)_B$ subgroups of the 
$SO(4)=SU(2)_A\times SU(2)_B$ $R$-symmetry group and are indicated by 
underlined and overlined indices, respectively. The same multiplet includes also 
$\bar q_{(I)\bar A}$ and $\bar \psi_{(I)}^{\ul B}$.
The field content of the twisted hypermultiplets is obtained by 
exchanging underlined and overlined indices.

The CS level of the $I$\textsuperscript{th} node, $k_I$, is fixed by the condition
\beq
k_I=\frac{k}{2}(s_I-s_{I+1})\,,
\qquad 
s_I=\pm1\,,
\eeq
where $s_I=1$ for a hypermultiplet at the $I$\textsuperscript{th} link and $s_I=-1$ 
for a twisted hypermultiplet.%
\footnote{We adopt the convention that in quiver diagrams $I$ increases from left to right.}
With these conditions, it is apparent that $k_I\in\{k,-k,0\}$.

\subsection{The Wilson loops}
\label{sec:inro-lines}

The theories we study have an $SO(4)=SU(2)_A\times SU(2)_B$ $R$-symmetry. We 
shall look for Wilson loops which preserve the $SU(2)_A$ subgroup. There will 
be other Wilson loops which preserve the $SU(2)_B$ subgroup, of course. Those can 
be studied by replacing the theory with another one where all hypermultiplets are 
exchanged with twisted hypermultiplets and vice-versa.

To be specific, we choose to preserve the supersymmetries generated by the four 
parameters (see the supersymmetry transformations in Appendix~\ref{app:not})
\beq\label{ta}
\xi_{\bar A \ul 1}^+,\quad
\xi_{\bar A \ul 2}^-\,,
\qquad
\bar A=\bar 1,\bar2\,.
\eeq
The Wilson loops preserving those supercharges are straight lines in Euclidean space 
and will also preserve four superconformal generators. In Section~\ref{sec:circ} we 
write down circular loops, which preserve eight linear combinations of the 
Poincar\'e and conformal supersymmetries, but will also preserve $SU(2)_A$.

An important point about the supersymmetries in \eqn{ta} is that there is a pairwise 
symmetry under exchange of chirality $\pm$ and $SU(2)_B$ index $\ul1$, $\ul2$. The 
construction of the Wilson loops below mirrors that in ABJM theory and includes 
fermions from the hypermultiplet, which carry the same type of indices. Hence the choice of 
fermionic coupling in the Wilson loops breaks this pairwise symmetry and consequently 
we find a pairwise degeneracy in all our constructions. We refer to those below as the 
``$\psi_1$-loops'' and ``$\psi_2$-loops'', reflecting the $SU(2)_B$ label of the fermionic coupling. 

There are several different cases of loops, the details of which are presented in the next 
section, but most possess this degeneracy. We do not see a trace of this degeneracy in 
M-theory and we expect it to be lifted by quantum corrections. We discuss this 
in some more detail in the discussion section.

The paper is organized as follows.
In the next section we present the construction of the $1/2$ BPS Wilson loops. 
We start with the case of segments of the quiver with alternating $k, -k$ CS levels. 
We then consider linear quivers, by removing a hypermultiplet or twisted hypermultiplet 
from a circular quiver. Finally, we consider the cases of segments of the quiver with 
vanishing CS levels.
In Section~\ref{sec:circ} we study the case of the circular Wilson loop. 
In Section~\ref{sec:loc} we discuss how to calculate those Wilson loops using localization. 
The main point is that all those Wilson loops for quivers with alternating CS levels 
are classically cohomologically equivalent to 
$1/4$ BPS loops, which can be evaluated in the matrix models of these theories.
We expect, though, that this analysis receives quantum corrections and only a certain linear 
combination of those Wilson loops will be in fact quantum mechanically equivalent to the 
$1/4$ BPS loops. 
For segments of the quiver comprising successive hypermultiplets, we have been able 
to show only for one of our proposed BPS loops that it is cohomologically equivalent 
to a $1/4$ BPS loop. The situation with the other possible BPS Wilson loops supported 
on this part of the quiver the situation is not clear. 
In Section~\ref{sec:M-theory} we discuss the M2-brane duals of these Wilson loops and 
comment on their degeneracies. We finally conclude with a discussion of the many remaining 
questions left open. The notations and some technicalities are relegated to appendices.

\paragraph{}

During the course of our work, a manuscript addressing the same question has appeared 
\cite{wu_zhang}. 
This prompted us to present the rich class of observables we have found, leaving their 
further study to the future. The Wilson loop found in \cite{wu_zhang} is one of those presented 
below for the particular theories which are orbifolds of ABJM. To be specific, it is the loop coupling to 
$\psi_1$ and in a representation of {\em all} of the nodes of the quiver.

\section{Infinite straight line}
\label{sec:sl}

Wilson loops are traditionally defined as the holonomy of the gauge connection. 
It was however found in \cite{drukker} that the $1/2$ BPS Wilson loops in ABJ(M) 
theory must be expressed in terms of 
the holonomy of a \emph{superconnection} $\cL$, 
with modified connections for the two $U(N_1)$ and $U(N_2)$ vector multiplets 
on the diagonal blocks and bifundamental Fermi fields in the off-diagonal blocks. 
The Wilson loop is then
\beq\label{wl}
\cW=\tr_{\cR} P \exp\left(-i\int d\tau \cL(\tau)\right),
\eeq
where $\cR$ is a representation of the supergroup $U(N_1|N_2)$.

The supersymmetry variation of the superconnection does not vanish, rather it is a 
total differential. In \cite{cardinali} this was expressed in terms of a 
supercovariant derivative of a supermatrix valued in the superalgebra, \emph{i.e.},
\beq\label{cov_der}
\delta\cL
=\mathfrak{D}_\tau G
\equiv \partial_\tau G-i\{\cL,G]\,.
\eeq
This is enough to guarantee that the Wilson loop is invariant under the corresponding 
supersymmetries. One must furthermore ensure that the trace is such that the boundary 
terms from integrating this term cancel each other. Our construction below of $1/2$ BPS loops 
in $\cN=4$ theories will be based on similar principles.

In this section, we start by considering an infinite spacelike line in the $x^1$-direction in $\bR^{3}$ parametrized by
\beq
x^1=\tau\,,\qquad
x^2=x^3=0\,.
\eeq

\subsection{Alternating CS levels}
\label{sec:k-k}

The theories we consider have a number of vector multiplets coupled by 
$p$ hypermultiplets and $q$ twisted hypermultiplets forming a circular or linear quiver. We study 
different possible sections of the quiver and find the Wilson loops supported on the 
relevant nodes.

We start by considering a segment of the quiver of the form
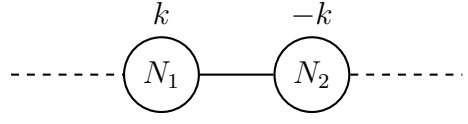
\begin{center}
\begin{tikzpicture}[every node/.style={circle,draw},thick]
\node(NL) at (2,0){$N_1$};
\node(NR) at (4,0){$N_2$};
\draw[dashed](0,0)--(NL);
\draw[solid](NL)--(NR);
\draw[dashed](NR)--(6,0);
\begin{scope}[nodes = {draw = none,above = 12pt}]
\node at (NL) {$k$};
\end{scope}
\begin{scope}[nodes = {draw = none,above = 8pt}]
\node at (NR) {$-k$};
\end{scope}
\end{tikzpicture}\,,
\end{center}
where a solid link corresponds to a hypermultiplet and a dashed link to a twisted hypermultiplet.
Each node represents a $U(N_I)$ vector multiplet with Chern-Simons level $k_I$ 
indicated above.

We begin by considering the variation of the gauge connection of the first node. 
Before proceeding, we define following \cite{im_aux}, the useful bilinears
\bal\label{mm}
&\nu_{(I)}=q_{(I)}^{\bar A} \bar q_{(I)\bar A},
&&\tilde\nu_{(I)}=\bar q_{(I)\bar A} q_{(I)}^{\bar A}\,,
\\
&(\mu_{(I)})^{\bar A}{}_{\bar B}
=q_{(I)}^{\bar A} \bar q_{(I)\bar B}
-\frac{1}{2}\delta^{\bar A}{}_{\bar B} (\nu_{(I)})^{\bar C}{}_{\bar C},
&&
(\tilde\mu_{(I)})^{\bar A}{}_{\bar B}
=\bar q_{(I)\bar B} q_{(I)}^{\bar A}
-\frac{1}{2}\delta^{\bar A}{}_{\bar B} (\tilde\nu_{(I)})^{\bar C}{}_{\bar C}\,,
\\
&j_{(I)}^{\bar A\ul B a}
=\sqrt{2}q_{(I)}^{\bar A}\bar\psi_{(I)}^{\ul B a}
-\sqrt{2}\epsilon^{\bar A\bar C}\epsilon^{\ul B\ul D}\psi_{(I)\ul D}^a \bar 
q_{(I)\bar C}\,,
\ \ &&
\tilde j_{(I)}^{\bar A\ul B a}
=\sqrt{2}\bar\psi_{(I)}^{\ul B a} q_{(I)}^{\bar A}
-\sqrt{2}\epsilon^{\bar A\bar C}\epsilon^{\ul B\ul D}\bar q_{(I)\bar C}\psi_{(I)
\ul D}^a\,.
\eal
The currents $j$ and $\tilde j$ are descendents of the moment maps $\mu$ and $
\tilde\mu$.

The variation with supersymmetry parameters \eqn{ta} is (see \app{not})
\beq
\label{deltaA11}
\delta A_{(1)1}=
\frac{1}{k} \xi_{\bar A \ul 1}^+ \left(j_{(1)+}^{\bar A \ul 1}-\tilde j_{(0)+}
^{\ul 1 \bar A}\right)
-\frac{1}{k} \xi_{\bar A \ul 2}^- \left(j_{(1)-}^{\bar A \ul 2}-\tilde j_{(0)-}
^{\ul 2 \bar A}\right).
\eeq

In CS-matter theories, it is natural to allow for a bilinear of the scalars in the connection 
\cite{yin, Chen:2008bp, plefka, Rey:2008bh, drukker}. 
The variation of the moment map \eqn{mm} 
associated to the twisted hypermultiplet on the left is
\beq
\delta (\tilde\mu_{(0)})^{\ul1}{}_{\ul1}
=\frac i2\xi_{\bar A\ul1}^+\tilde j_{(0)+}^{\bar A\ul1}-\frac i2\xi_{\bar A\ul2}
^-\tilde j_{(0)-}^{\bar A\ul1}\,.
\eeq
We can eliminate all the terms in the variation of $A_{(1)}$ \eqn{deltaA11} 
that depend on the twisted hypermultiplets by taking the linear combination
\beq\label{bos}
\delta\left[A_{(1)1}-\frac{2i}{k}(\tilde\mu_{(0)})^{\ul1}{}_{\ul1}\right]
=\frac{1}{k} \xi_{\bar A \ul 1}^+ j_{(1)+}^{\bar A \ul 1}
-\frac{1}{k} \xi_{\bar A \ul 2}^- j_{(1)-}^{\bar A \ul 2}\,.
\eeq
Since $\tilde\mu$ is traceless, this is the same as adding 
$\frac{2i}{k}(\tilde\mu_{(0)})^{\ul2}{}_{\ul2}$.

We can also include terms from the scalars in the untwisted hypermultiplets. 
A term proportional to their moment map will lead to a connection invariant 
under half of the supercharges in \eqn{ta}, that is a $1/4$ BPS Wilson loop. 
But our choice of of supersymmetries \eqn{ta} distinguish between the twisted 
and untwisted fields. So for the untwisted fields consider the variation of $\nu$
\beq
\label{nu}
\delta \nu_{(1)}
=i\sqrt{2}\left(\epsilon^{\bar A\bar B}\epsilon^{\ul1\ul2}\xi^-_{\bar B \ul 2} 
\psi_{(1)\ul 1-} \bar q_{\bar A}
+\epsilon^{\bar A\bar B}\epsilon^{\ul2\ul1}\xi^+_{\bar B \ul 1} \psi_{(1)\ul 2+} 
\bar q_{\bar A}
+\xi_{\bar A \ul 1}^+ q^{\bar A}_{(1)} \bar\psi_{(1)+}^{\ul1}
+\xi_{\bar A \ul 2}^- q^{\bar A}_{(1)} \bar\psi_{(1)-}^{\ul2}\right).
\eeq
This does not package nicely in terms of the currents, so let us also 
expand \eqn{bos} in terms of the component fields
\bal\label{A_exp}
&\delta\left[A_{(1)1}-\frac{2i}{k}(\tilde\mu_{(0)})^{\ul1}{}_{\ul1}\right]
\\&\qquad
=\frac{\sqrt{2}}{k}\left(\xi_{\bar A \ul 1}^+ q^{\bar A}_{(1)} \bar\psi_{(1)+}
^{\ul1}
-\xi_{\bar A \ul 2}^- q^{\bar A}_{(1)} \bar\psi_{(1)-}^{\ul2}
-\epsilon^{\bar A\bar B}\epsilon^{\ul2\ul1}\xi^+_{\bar B \ul1} \psi_{(1)\ul2+} 
\bar q_{(1)\bar A}
+\epsilon^{\bar A\bar B}\epsilon^{\ul1\ul2}\xi^-_{\bar B \ul 2} \psi_{(1)\ul1-} 
\bar q_{(1)\bar A}\right).
\eal
We see that we have all the same terms in those two expressions, but with different 
signs. We can therefore add or subtract $\nu_{(1)}$ from the gauge connection and reduce the 
variation to just two terms
\begin{subequations}
\begin{align}
\label{mod_c}
\delta\left[A_{(1)1}-\frac{2i}{k}(\tilde\mu_{(0)})^{\ul1}{}_{\ul1}-\frac{i}{k}\nu_{(1)}\right]
&=\frac{2\sqrt{2}}{k}\left[\xi_{\bar A\ul1}^+ q_{(1)}^{\bar A} \bar\psi_{(1)+}^{\ul1}
+\epsilon^{\bar A\bar B}\epsilon^{\ul1\ul2}\xi^-_{\bar B\ul2} \psi_{(1)\ul1-} 
\bar q_{(1)\bar A}\right],
\\
\label{mod_c2}
\delta\left[A_{(1)1}-\frac{2i}{k}(\tilde\mu_{(0)})^{\ul1}{}_{\ul1}+\frac{i}{k}\nu_{(1)}\right]
&=-\frac{2\sqrt{2}}{k}\left[\xi_{\bar A\ul2}^- q_{(1)}^{\bar A} \bar\psi_{(1)-}^{\ul2}
+\epsilon^{\bar A\bar B}\epsilon^{\ul2\ul1}\xi^+_{\bar B\ul1} \psi_{(1)\ul2+} 
\bar q_{(1)\bar A}\right].
\end{align}
\end{subequations}

In each case we find a non-vanishing variation, which we will have to cancel by considering 
a superconnection, with fermionic couplings either $\bar\psi_{(1)+}^{\ul1}$ and 
$\psi_{(1)\ul1-}$ for the first sign choice, or $\bar\psi_{(1)-}^{\ul2}$ and $\psi_{(1)\ul2+}$ 
in the second case. 
We refer in the following to the two respective loops as ``$\psi_1$-loops'' and ``$\psi_2$-loops''. 

\subsubsection{The \texorpdfstring{$\psi_1$}{psi1}-loop}
\label{sec:a1}

Thus, following \cite{drukker,cardinali}, we try introducing a super-connection 
whose top left block is given by \eqn{mod_c}.
On dimensional and Gra{\ss}mann odd/even grounds, it can be seen that 
the (1,2) and (2,1) component of the superconnection will be of the form, $\bar c^{\ul A}_a 
\psi_{(1)\ul A}^a$ and $c_{\ul A}^a \bar\psi_{(1)a}^{\ul A}$, where the spinor 
couplings $\bar c^{\ul A}_a$ and $c_{\ul A}^a$ are Gra{\ss}mann even.
The supersymmetry conditions \eqn{cov_der} tell us that we must write the variation of 
this entry with respect to the supersymmetries as a covariant derivative 
with respect to some specified modified bosonic connections.
By assumption, we firstly consider taking the modified bosonic connection of the 
first node is given by \eqn{mod_c}, and we fix the form of the second 
connection such that the conditions \eqn{cov_der} are satisfied.

As such, we must choose the fermionic couplings such that the covariant 
derivative in the variation of the fermions is projected along the $x^1$ direction.
This requires $c_{\ul1}^-=c_{\ul2}^+=\bar c^{\ul1}_-=\bar c^{\ul2}_+=0$. 
Motivated by the lack of an appearance of $\psi_2$ and $\bar\psi^2$ in 
\eqn{mod_c} in the context of the supersymmetry conditions \eqn{cov_der}, let us further 
assume that $c_{\ul2}^-=\bar c^{\ul2}_-=0$.
We thus write the superconnection as
\beq
\cL^{\psi_1}=
\begin{pmatrix}
A_{(1)1}-\frac{2i}{k}(\tilde\mu_{(0)})^{\ul1}{}_{\ul1}-\frac{i}{k}\nu_{(1)}\quad 
& \bar c \psi_{(1)\ul1}^+\\
c \bar\psi_{(1)+}^{\ul1} & \star
\end{pmatrix},
\eeq
where $c$, $\bar c$ and `$\star$' remain to be fixed.

As the variation of the bosonic connection contains no derivatives, the 
supermatrix $G$ of \eqn{cov_der} is of the form
\beq
G^{\psi_1}=
\begin{pmatrix}
0 & g\\
\bar g & \star
\end{pmatrix}.
\eeq
Towards this, consider
\bal
\delta\psi_{(1)\ul1}^+=&\,
\sqrt2 (\slashed D)^+{}_+ q_{(1)}^{\bar A} \, \xi_{\bar A\ul1}^+
-\frac{\sqrt2}{k} \xi_{\bar A\ul1}^+
\left(\nu_{(1)} q_{(1)}^{\bar A}
-q_{(1)}^{\bar A} \tilde\nu_{(1)}\right)\\
&{}-\frac{2\sqrt2}{k} \xi_{\bar A\ul1}^+ (\tilde\mu_{(0)})^{\ul 1}_{\ \ul 1} 
q_{(1)}^{\bar A}
+\frac{2\sqrt2}{k} \xi_{\bar A \ul1}^+ q_{(1)}^{\bar A} (\mu_{(2)})^{\ul1}_{\ \ul 
1}.
\eal
Thus, we may write
\beq
\bar c\,\delta\psi_{(1)\ul1}^+=\cD_\tau g,
\eeq
with
\bal\label{cd1}
&g\equiv
\sqrt{2}(\bar c^{\ul1})_+q_{(1)}^{\bar A}(\xi_{\bar A\ul1})^+\,,
\\
&\cD_\tau g\equiv 
D_\tau g+i\left(\frac{i}{k}\nu_{(1)}+\frac{2i}{k}
(\tilde\mu_{(0)})^{\ul1}{}_{\ul1}\right)g
-ig\left(\frac{i}{k}\tilde\nu_{(1)}+\frac{2i}{k}
(\mu_{(2)})^{\ul1}{}_{\ul1}\right),
\\
&D_\tau g\equiv \partial_\tau g-iA_{(1)1}g+igA_{(2)1}\,,
\eal
in agreement with the modified connection for the loop in \eqn{mod_c}.
We are now furthermore find that the bottom right block has be be 
$A_{(2)}-\frac{2i}{k}(\mu_{(2)})^{\ul1}{}_{\ul1}-\frac{i}{k}\tilde\nu_{(1)}$. 

Similarly, we may use the variation of the bottom left block
\bal
\delta\bar\psi_{(1)+}^{\ul1}
=&\,\sqrt{2}(\slashed D)_{+-}\bar q^{\bar A}_{(1)}\epsilon^{\bar A\bar B}
\epsilon^{\ul1\ul2}\xi^-_{\bar B\ul2}
+\frac{\sqrt{2}}{k}\epsilon^{\bar A\bar B}\epsilon^{\ul1\ul2}\xi_{\bar B\ul 2+}
\left[\tilde\nu_{(1)} \bar q_{(1)\bar A}
-\bar q_{(1)\bar A} \nu_{(1)}\right]\\
&-\frac{2\sqrt{2}}{k}\epsilon^{\bar A\bar B}\epsilon^{\ul1\ul2}\xi_{\bar B\ul 2+} 
\bar q_{(1)\bar A} (\tilde\mu_{(0)})^{\ul 1}{}_{\ul 1}
+\frac{2\sqrt{2}}{k}\epsilon^{\bar A\bar B}\epsilon^{\ul1\ul2}\xi_{\bar B\ul 2+} 
(\mu_{(2)})^{\ul 1}{}_{\ul 1} \bar q_{(1)\bar A},
\eal
to write
\beq
c\,\delta\bar\psi_{(1)+}^{\ul1}=\cD_\tau \bar g,
\eeq
where
\beq
\bar g\equiv-
\sqrt{2}\,c\,\bar q^{\bar A}_{(1)}\epsilon^{\bar A\bar B}\epsilon^{\ul1\ul2}
\xi_{\bar B\ul2+},
\eeq
and the modified bosonic connections agree with \eqn{cd1}.
It now remains to check the variations of the bosonic connections satisfy the 
conditions coming from \eqn{cov_der}, \emph{i.e.},
\begin{subequations}
\begin{align}
&\delta
\left[A_{(1)1}-\frac{2i}{k}(\tilde\mu_{(0)})^{\ul1}{}_{\ul1}-\frac{i}{k}\nu_{(1)}\right]
=-i\bar c\,(\psi_{(1)\ul1})^+\bar g
+igc\,(\bar\psi_{(1)}^{\ul1})_+,\\
&\delta\left[A_{(2)1}-\frac{2i}{k}(\mu_{(2)})^{\ul1}{}_{\ul1}-\frac{i}{k}\tilde\nu_{(1)}\right]
=-ic\,(\bar\psi_{(1)}^{\ul1})_+g
+i\bar g\bar c\,(\psi_{(1)\ul1})^+.
\end{align}
\end{subequations}
A simple calculation shows that these conditions hold, provided $c\bar c=-\frac{2i}{k}$. 
The value of $c$ itself is immaterial, as it always will appear in the combination 
$c\bar c$ in the trace of the superconnection. 
We choose therefore $c=\bar c=\frac{1-i}{\sqrt k}$. 
In summary, we find that the loop with super-connection
\beq
\label{lpsi1}
\cL^{\psi_1}=
\begin{pmatrix}
A_{(1)1}-\frac{2i}{k}(\tilde\mu_{(0)})^{\ul1}{}_{\ul1}-\frac{i}{k}\nu_{(1)}
&\frac{1-i}{\sqrt k}\,\psi_{(1)\ul1}^+\\
\frac{1-i}{\sqrt k}\,\bar\psi_{(1)+}^{\ul1}
&A_{(2)1}-\frac{2i}{k}(\mu_{(2)})^{\ul1}{}_{\ul1}-\frac{i}{k}\tilde\nu_{(1)}
\end{pmatrix}
\eeq
preserves the supersymmetries \eqn{ta}.

\subsubsection{The \texorpdfstring{$\psi_2$}{psi2}-loop}
\label{sec:p2}

We now consider the second modified bosonic connection \eqn{mod_c2} 
$A_{(1)1}-\frac{2i}{k}(\tilde\mu_{(0)})^{\ul1}{}_{\ul1}+\frac{i}{k}\nu_{(1)}$ 
and proceed similarly to the previous section.

We find that this works with
\beq
\bar g\equiv
\sqrt{2}\,c\,\bar q_{(1)\bar A}\epsilon^{\bar A\bar B}\epsilon^{\ul2\ul1}
\xi_{\bar B\ul1-},
\eeq
and the super-connection of the $\psi_2$-loop is given by
\beq\label{lpsi2}
\cL^{\psi_2}=
\begin{pmatrix}
A_{(1)1}-\frac{2i}{k}(\tilde\mu_{(0)})^{\ul1}{}_{\ul1}+\frac{i}{k}\nu_{(1)}
&\frac{1-i}{\sqrt k}\,\psi_{(1)\ul2}^-\\
\frac{1-i}{\sqrt k}\,\bar\psi_{(1)-}^{\ul2}
&A_{(2)1}-\frac{2i}{k}(\mu_{(2)})^{\ul1}{}_{\ul1}+\frac{i}{k}\tilde\nu_{(1)}
\end{pmatrix}.
\eeq

Indeed using
\bal
\delta\psi_{(1)\ul2}^-=&\,
\sqrt2 (\slashed D)^-{}_- q_{(1)}^{\bar A} \, \xi_{\bar A\ul2}^-
-\frac{\sqrt2}{k} \xi_{\bar A\ul2}^-
\left(\nu_{(1)} q_{(1)}^{\bar A}
-q_{(1)}^{\bar A} \tilde\nu_{(1)}\right)\\
&{}-\frac{2\sqrt2}{k} \xi_{\bar A\ul2}^- (\tilde\mu_{(0)})^{\ul 2}_{\ \ul 2} 
q_{(1)}^{\bar A}
+\frac{2\sqrt2}{k} \xi_{\bar A \ul2}^- q_{(1)}^{\bar A} (\mu_{(2)})^{\ul2}_{\ \ul 
2}\,,
\eal
and the variation of the bosonic connections we find that 
the supermatrix appearing in the covariant derivative \eqn{cov_der} is
\beq
G^{\psi_2}=
\begin{pmatrix}
0 & -\frac{\sqrt{2}(1-i)}{\sqrt k}\,q_{(1)}^{\bar A} \, \xi_{\bar A\ul2}^-
\\
\frac{\sqrt{2}(1-i)}{\sqrt k}\,\bar q_{(1)\bar A}\epsilon^{\bar A\bar B}\epsilon^{\ul2\ul1}\xi_{\bar B\ul1-}
& 0
\end{pmatrix}.
\eeq

We find for each hypermultiplet in a quiver connecting nodes with CS levels 
$k$ to $-k$ a pair of superconnections. Out of each of them we can construct a Wilson 
loop in an arbitrary representation of the supergroup $U(N_1|N_2)$. We in fact expect 
to find only one Wilson loop for each representation of such a pair. We comment about this 
in the discussion section.

%%%%%%%%%%%%

\subsection{Linear quivers}
\label{sec:line}

Before proceeding to the case of quivers with nodes with vanishing CS levels, let us 
comment on linear quivers. We can construct any linear quiver by starting from a circular 
quiver and removing a hypermultiplet or a twisted hypermultiplet (or a pair and the intermediate 
vector multiplet). The relevant loops are gotten from those constructed above by erasing 
the coupling to the removed fields.

We start from the segment of the quiver considered above
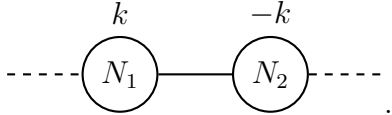
\begin{center}
\begin{tikzpicture}[every node/.style={circle,draw},thick]
\node(NL) at (2,0){$N_1$};
\node(NR) at (4,0){$N_2$};
\draw[dashed](0.5,0)--(NL);
\draw[solid](NL)--(NR);
\draw[dashed](NR)--(5.5,0);
\begin{scope}[nodes = {draw = none,above = 12pt}]
\node at (NL) {$k$};
\end{scope}
\begin{scope}[nodes = {draw = none,above = 8pt}]
\node at (NR) {$-k$};
\end{scope}
\end{tikzpicture}.
\end{center}
The loops we constructed couple to scalars and fermions of the hypermultiplet between 
the nodes and to the scalars of the two adjacent twisted hypermultiplets. 

Consider removing the twisted hypermultiplet to the right of the second node. Erasing it from 
the $\psi_1$-loop \eqn{lpsi1} gives
\beq
\cL^{\psi_1}=
\begin{pmatrix}
A_{(1)1}-\frac{2i}{k}(\tilde\mu_{(0)})^{\ul1}{}_{\ul1}-\frac{i}{k}\nu_{(1)}
&\frac{1-i}{\sqrt k}\,\psi_{(1)\ul1}^+\\
\frac{1-i}{\sqrt k}\,\bar\psi_{(1)+}^{\ul1}
&A_{(2)1}-\frac{i}{k}\tilde\nu_{(1)}
\end{pmatrix},
\eeq
and it is easy to check that this is indeed is $1/2$ BPS, with the appropriate supersymmetry 
transformations of the linear quiver in \app{not}. 
Similarly, the $\psi_2$-loop \eqn{lpsi2} becomes
\beq
\cL^{\psi_2}=
\begin{pmatrix}
A_{(1)1}-\frac{2i}{k}(\tilde\mu_{(0)})^{\ul1}{}_{\ul1}+\frac{i}{k}\nu_{(1)}
&\frac{1-i}{\sqrt k}\,\psi_{(1)\ul2}^-\\
\frac{1-i}{\sqrt k}\,\bar\psi_{(1)-}^{\ul2}
&A_{(2)1}+\frac{i}{k}\tilde\nu_{(1)}
\end{pmatrix}.
\eeq
Likewise we can remove the twisted hypermultiplet from the left which will remove the 
coupling to the $\tilde \mu_{(0)}$ moment map from the above.

If we instead remove the hypermultiplet connecting the $N_1$ and $N_2$ nodes we 
expect to lose the structure of the superconnection. Indeed, in that case we find two 
independent $1/2$ BPS Wilson loops, each with a gauge connection and a coupling 
to the twisted moment maps
\beq
\label{1/4=1/2}
\cL_{(1)}=A_{(1)1}-\frac{2i}{k}(\tilde\mu_{(0)})^{\ul1}{}_{\ul1}\,,
\qquad
\cL_{(2)}=A_{(2)1}-\frac{2i}{k}(\mu_{(2)})^{\ul1}{}_{\ul1}\,.
\eeq
Note that both $\psi_1$ and $\psi_2$-loops give the same 
loops via this process, so we lose that two-fold degeneracy. 
Also, these loops are essentially the same as the usual $1/4$ BPS loops one gets 
from a modified connection with the scalar $\varphi$ of the vector multiplet. In this 
formulation this scalar has been integrated out giving rise to the couplings to $\mu$. 
Those loops are $1/2$ BPS in theories with $\cN=2$ supersymmetry and in this 
particular case get enhanced to preserve 4 supercharges, rather than 2.

A similar story holds when considering linear quivers that end on nodes with vanishing 
CS levels. The case with vanishing levels inside a quiver are studied in the following 
two sections. Again, one can remove a hypermultiplet or twisted hypermultiplet to 
open up the quiver and the Wilson loops with those fields removed would remain $1/2$ 
BPS. We do not list all the examples or repeat the algebra for all of them.

\subsection{Vanishing CS levels: Repeated untwisted hypermultiplets}
\label{sec:k0-k}

We have thus far considered Chern-Simons-matter quivers with alternating $\pm k$ 
Chern-Simons levels. Here we study the Wilson loops in the case with extra 
nodes of vanishing CS levels \cite{gaiotto,im_aux} that preserve $\cN=4$ supersymmetry.

Let us consider the quiver with a node with vanishing CS coupling to a {\em pair} of 
untwisted hypermultiplets
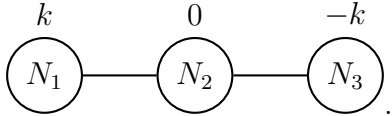
\begin{center}
\begin{tikzpicture}[every node/.style={circle,draw},thick]
\node(NL) at (2,0){$N_1$};
\node(NM) at (4,0){$N_2$};
\node(NR) at (6,0){$N_3$};
\draw[solid](NL)--(NM);
\draw[solid](NM)--(NR);
\begin{scope}[nodes = {draw = none,above = 12pt}]
\node at (NL) {$k$};
\node at (NM) {0};
\end{scope}
\begin{scope}[nodes = {draw = none,above = 8pt}]
\node at (NR) {$-k$};
\end{scope}
\end{tikzpicture}.
\end{center}
We have removed for convenience the twisted hypermultiplets from the left and the right, 
but they can be incorporated by adding the moment maps to the gauge connection. 

We start by mimicking the construction of superconnections in the alternating 
levels case and find several different possible connections involving pairs of nodes. 
Then in Section~\ref{sec:4x4} 
we find a new structure of a superconnection which couples to all the different fields 
of the {\em three} nodes. We expect this structure to be more fundamental than 
the ones involving only pairs of nodes.

\subsubsection{The \texorpdfstring{$\psi_1$}{psi1}-loop for the first link}
\label{sec:k0l1}

We begin by constructing the $\psi_1$-loop for the first link.
The analysis proceeds as in \sect{a1}: We take the same connection for the 
first node, since it couples to the same fields as previously (only that now, since 
we don't have twisted hypermultiplets, we don't need to include a $\tilde\mu_{(0)}$ contribution). 
We then continue to study the fermions and the connection on the second node. 

Requiring that the variation of the bosonic connection in the first node is a commutator 
with $\psi_{(1)}$ gives%
\footnote{
The `$(1)$' subscript for $\cL$ indicates that we are attempting to 
construct a $U(N_1|N_2)$ valued connection.}
\beq
\cL^{\psi_1}_{(1)}=
\begin{pmatrix}
A_{(1)1}-\frac{i}{k}\nu_{(1)}\quad
&\bar c\,\psi_{(1)\ul1}^+\\
c\,\bar\psi_{(1)+}^{\ul1}
&\star
\end{pmatrix},
\eeq
where it remains to fix `$\star$' and the fermionic couplings $c$ and $\bar c$. 

We fix the form of the bosonic connection of the 
second node by studying the variation of the fermionic terms
\bal
\delta\psi_{(1)\ul 1}^+=&\,
\sqrt{2}D_\tau q^{\bar A}_{(1)}\xi_{\bar A\ul 1}^+
-\frac{\sqrt{2}}{k}\xi_{\bar A\ul 1}^+
\left[\nu_{(1)} q^{\bar A}_{(1)}
-q^{\bar A}_{(1)} \tilde\nu_{(1)}\right]\\
&-\frac{2\sqrt{2}}{k}\xi_{\bar A\ul 1}^+ 
(\tilde\mu_{(0)})^{\ul 1}{}_{\ul 1} q^{\bar A}_{(1)}
-\sqrt{2}\xi_{\bar A\ul1}^+ q^{\bar A}_{(1)} (\varphi_{(2)})^{\ul 1}{}_{\ul 1}.
\eal
If we require that the variation of the fermion is a total derivative this fixes 
the second connection to 
$A_{(2)1}+i(\varphi_{(2)})^{\ul1}{}_{\ul1}-\frac{i}{k}\tilde\nu_{(1)}$.

Furthermore, it fixes
\beq
g_{(1)}=
\sqrt2 \bar c q^{\bar A}_{(1)} \xi_{\bar A\ul1}^+.
\eeq
It is also in agreement with the bosonic connection of the first node given above.
The variation of the conjugate fermion suggests the same bosonic connections as 
well as
\beq
\bar g_{(1)}=
-\sqrt2 c\,\bar q_{(1)\bar A}\epsilon^{\bar A\bar B}\epsilon^{\ul1\ul2}\xi_{\bar 
B\ul2+},
\eeq
\emph{cf.} the calculation in \sect{a1}.
This suggests that the form of the supermatrix $G_{(1)}$ is identical to that of 
the alternating level case.

We see that if we take
\beq
\label{k0-kpsi11}
\cL^{\psi_1}_{(1)}=
\begin{pmatrix}
A_{(1)1}-\frac{i}{k}\nu_{(1)}\quad
&\frac{1-i}{\sqrt k}\,\psi_{(1)\ul1}^+\\
\frac{1-i}{\sqrt k}\,\bar\psi_{(1)+}^{\ul1}
&A_{(2)1}+i(\varphi_{(2)})^{\ul1}{}_{\ul1}-\frac{i}{k}\tilde\nu_{(1)}
\end{pmatrix},
\eeq
then we verified that the variation of all but the bottom right block are total covariant 
superderivatives. The variation of that block is
\begin{align}
\label{split}
&\delta
\left[
A_{(2)1}+i(\varphi_{(2)})^{\ul1}{}_{\ul1}-\frac{i}{k}\tilde\nu_{(1)}\right]
\\
&=\frac{2\sqrt{2}}{k}\left[
\xi_{\bar A\ul1}^+\bar\psi_{(1)+}^{\ul 1} q_{(1)}^{\bar A}
+\xi^{\bar A\ul 1-} \bar q_{(1)\bar A} \psi_{(1)\ul 1-}\right]
+\frac{1}{2k}\left(\xi_{\bar A\ul1}^+\left[
j_{(2)+}^{\bar A\ul1} - \tilde j_{(1)+}^{\bar A\ul1}\right]
-\xi_{\bar A\ul2}^-\left[
j_{(2)-}^{\bar A\ul2} - \tilde j_{(1)-}^{\bar A\ul2}
\right]\right).
\nonumber
\end{align}
The first two terms are exactly as for the variation of the 
bosonic connection of the second node in \sect{a1}. Thus we can write 
\beq\label{delj}
\delta\cL_{(1)}^{\psi_1}\equiv\mathfrak D_{(1)\tau}^{\psi_1} G_{(1)}^{\psi_1} + \Delta J
\begin{pmatrix}
0 & 0\\
0 & 1
\end{pmatrix},
\qquad
\Delta J=\frac{1}{2k}
\left(\xi_{\bar A\ul1}^+\left[
j_{(2)+}^{\bar A\ul1} - \tilde j_{(1)+}^{\bar A\ul1}\right]
- \xi_{\bar A\ul2}^-\left[
j_{(2)-}^{\bar A\ul2} - \tilde j_{(1)-}^{\bar A\ul2}\right]\right)\,,
\eeq
where $\mathfrak D_{(1)\tau}^{\psi_1}$ is a supercovariant derivative with respect 
the superconnection $\cL^{\psi_1}_{(1)}$.

We shall come back to discuss the $\Delta J$ term in Section~\ref{sec:J}. For now 
we examine more possibilities for superconnections.

\subsubsection{The \texorpdfstring{$\psi_2$}{psi2}-loop for the first link}

We can also take the superconnection
\beq
\cL^{\psi_2}_{(1)}
=\begin{pmatrix}
A_{(1)1}+\frac{i}{k}\nu_{(1)}\quad
&\bar c\,\psi_{(1)\ul2}^-\\
c\,\bar\psi_{(1)-}^{\ul2}
&A_{(2)1}+i(\varphi_{(2)})^{\ul1}{}_{\ul1}+\frac{i}{k}\tilde\nu_{(1)}
\end{pmatrix},
\eeq
and find
\beq
\delta\cL_{(1)}^{\psi_2}
\equiv\mathfrak D^{\psi_2}_{(1)\tau} G_{(1)}^{\psi_2} + \Delta J
\begin{pmatrix}
0 & 0\\
0 & 1
\end{pmatrix}\,,
\eeq
where $G_{(1)}^{\psi_2}$ is the supermatrix for the alternating level case.

\subsubsection{Superconnections in the second link}

We can also consider superconnections involving the vector fields 
$A_{(2)}$ and $A_{(3)}$ and the hypermultiplet connecting them. By reflection 
it is clear that the resulting loops would have superconnections
\beq
\cL^{\psi_1}_{(1)}=
\begin{pmatrix}
A_{(2)1}+ i (\varphi_{(2)})^{\ul1}{}_{\ul1}-\frac{i}{k}\nu_{(2)} 
&\frac{1-i}{\sqrt k}\,\psi_{(2)\ul1}^+\\
\frac{1-i}{\sqrt k}\,\bar\psi_{(2)+}^{\ul1}
&A_{(3)1}-\frac{i}{k}\tilde\nu_{(2)}
\end{pmatrix}\,,
\eeq
satisfying\beq
\delta\cL_{(2)}^{\psi_1}=
\mathfrak D_{(2)\tau}^{\psi_1} G_{(2)}^{\psi_1} - \Delta J
\begin{pmatrix}
1 & 0\\
0 & 0
\end{pmatrix},
\eeq
where $G_{(2)}^{\psi_1}$ is the supermatrix found in the alternating level case.

Likewise
\beq
\cL^{\psi_2}_{(2)}=
\begin{pmatrix}
A_{2)1}+ i(\varphi_{(2)})^{\ul1}{}_{\ul1}+\frac{i}{k}\nu_{(2)} 
&\frac{1-i}{\sqrt{k}}\psi_{(2)\ul2}^-\\
\frac{1-i}{\sqrt{k}}\bar\psi_{(2)-}^{\ul2}
&A_{(3)1}+\frac{i}{k}\tilde\nu_{(2)}
\end{pmatrix},
\eeq
satisfies
\beq
\delta\cL_{(2)}^{\psi_1}=
\mathfrak D_{(2)\tau}^{\psi_2} G_{(2)}^{\psi_2} - \Delta J
\begin{pmatrix}
1 & 0\\
0 & 0
\end{pmatrix},
\eeq
where $G_{(2)}^{\psi_2}$ is the supermatrix found in the alternating level case.

\subsubsection{A $4\times4$ construction}
\label{sec:4x4}

Thus far we have found four superconnections involving pairs of vector fields and the relevant 
hypermultiplets. We can construct from them Wilson loops in arbitrary representations of 
$U(N_1|N_2)$ and $U(N_2|N_3)$. We can also combine pairs of them together, to form 
block diagonal superconnections with a $4\times4$ block structure, with the central 
node represented twice. Those will allow to write Wilson loops in arbitrary representations 
of $U(N_1+N_3|2N_2)$.

In fact, this $4\times4$ ansatz allows for a more general superconnection, which is not 
block diagonal. In addition to bosonic entries transforming in the adjoints of the individual nodes, 
and fermions in the bifundamentals of adjacent nodes, 
we have two bosonic entries transforming in bifundamentals of $U(N_1)\times U(N_3)$.
On dimensional and representation grounds, they have to take the form 
$\bar d_{\bar A\bar B}q_{(1)}^{\bar A}q_{(2)}^{\bar B}$ and 
$d^{\bar A\bar B}\bar q_{(2)\bar A}q_{(1)\bar B}$, respectively. We got the construction 
below to work with the antisymmetric couplings 
$\bar d_{\bar A\bar B}=\bar d\epsilon_{\bar A\bar B}$ 
and $d^{\bar A\bar B}=d\epsilon^{\bar A\bar B}$.

Starting from a general ansatz we have been able to show that the 
structure of the superconnection has to be either the block diagonal ones made of pairs 
of superconnections discussed above or the superconnection
\bal\label{4x4}
&\cL=
\begin{pmatrix}
A_{(1)1} 
&\ \bar c_{(1)}^{\ul1}\psi_{(1)\ul1}^+ \ 
&\ \bar c_{(1)}^{\ul2}\psi_{(1)\ul2}^-\ 
&\ \bar d\epsilon_{\bar A\bar B} q_{(1)}^{\bar A}q_{(2)}^{\bar B}\\
c_{(1)\ul1}\bar\psi_{(1)+}^{\ul1} 
& \cA_{(2)}
& 0 & \bar c_{(2)}^{\ul2} \psi_{(2)\ul2}^-\\
c_{(1)\ul2}\bar\psi_{(1)-}^{\ul2} & 0
& \cA_{(2)}^\prime
& \bar c_{(2)}^{\ul1}\psi_{(2)\ul1}^+\\
d\epsilon^{\bar A\bar B}\bar q_{(2)\bar A}q_{(1)\bar B}\ 
&\ c_{(2)\ul2}\bar\psi_{(2)-}^{\ul2}\ 
&\ c_{(2)\ul1}\bar\psi_{(2)+}^{\ul1}\ 
& A_{(3)1}
\end{pmatrix},
\eal
with $\cA_{(2)}\equiv A_{(2)1}+i(\varphi_{(2)})^{\ul1}{}_{\ul1} 
+\frac{i}{2k}\left(\nu_{(2)}-\tilde\nu_{(1)}\right)$ 
and $\cA_{(2)}^\prime\equiv A_{(2)1}+i(\varphi_{(2)})^{\ul1}{}_{\ul1}
-\frac{i}{2k}\left(\nu_{(2)}-\tilde\nu_{(1)}\right)$. 
The connection in the top left and bottom right corners involve only the gauge fields 
(and one could include the appropriate moment maps, if coupling to extra twisted hypermultiplets). 
It is rather interesting that in the case of alternating levels we had to augment 
all gauge connections by a coupling to $\nu$ with a coefficient $\pm i/k$. Indeed that 
is also what we found in the loops associated to the first and second link. But here we 
have the third option of not including this coupling for the external nodes.%
\footnote{We were not able to find BPS Wilson loops with any other 
values for the coupling to $\nu$.}

The supermatrix $G$ is given by
\beq
G=
\begin{pmatrix}
0 & g_{(1)\ul1} & g_{(1)\ul2} & 0\\
\bar g_{(1)}^{\ul1} & 0 & 0 & g_{(2)\ul2}\\
\bar g_{(1)}^{\ul2} & 0 & 0 & g_{(2)\ul1}\\
0 & \bar g_{(2)}^{\ul2} & \bar g_{(2)}^{\ul1} & 0
\end{pmatrix},
\eeq
where
\bal
g_{(1)\ul1}&=\sqrt{2}\bar c_{(1)}^{\ul1} q_{(1)}^{\bar A}\xi_{\bar A\ul1}^+\,,\quad
& g_{(1)\ul2}&=-\sqrt{2}\bar c_{(1)}^{\ul2} q_{(1)}^{\bar A}\xi_{\bar A\ul2}^-\,,\quad\\
g_{(2)\ul1}&=\sqrt{2}\bar c_{(2)}^{\ul1} q_{(2)}^{\bar A}\xi_{\bar A\ul1}^+\,,\quad
& g_{(2)\ul2}&=-\sqrt{2}\bar c_{(2)}^{\ul2} q_{(2)}^{\bar A}\xi_{\bar A\ul2}^-\,,\quad\\
\bar g_{(1)}^{\ul1}&=-\sqrt{2} c_{(1)\ul1} \bar q_{(1)\bar A}\epsilon^{\bar A\bar B}\epsilon^{\ul1\ul2}\xi_{\bar B\ul2+}\,,\quad
& \bar g_{(1)}^{\ul2}&=\sqrt{2} c_{(1)\ul2} \bar q_{(1)\bar A}\epsilon^{\bar A\bar B}\epsilon^{\ul2\ul1}\xi_{\bar B\ul1-}\,,\quad\\
\bar g_{(2)}^{\ul1}&=-\sqrt{2} c_{(2)\ul1} \bar q_{(2)\bar A}\epsilon^{\bar A\bar B}\epsilon^{\ul1\ul2}\xi_{\bar B\ul2+}\,,\quad
& \bar g_{(2)}^{\ul2}&=\sqrt{2} c_{(2)\ul2} \bar q_{(2)\bar A}\epsilon^{\bar A\bar B}\epsilon^{\ul2\ul1}\xi_{\bar B\ul1-}\,.\quad
\eal
We firstly consider the variations of the bosonic components, listing below the corresponding conditions which must be imposed for supersymmetry.
\bal\label{constraints}
& \tx{11 entry:} & c_{(1)}^{\ul1}\bar c_{(1)\ul1}&=c_{(1)}^{\ul2}\bar c_{(1)\ul2}=-\frac{i}{k}\,,\\
& \tx{44 entry:} & c_{(2)}^{\ul1}\bar c_{(2)\ul1}&=c_{(2)}^{\ul2}\bar c_{(2)\ul2}=-\frac{i}{k}\,,\\
& \tx{14 entry:} & -\bar c_{(1)}^{\ul1} \bar c_{(2)}^{\ul2}
&=\bar c_{(1)}^{\ul2} \bar c_{(2)}^{\ul1}=\bar d\,,\\
& \tx{41 entry:} & -c_{(1)\ul1}c_{(2)\ul2}&=c_{(1)\ul2}c_{(2)\ul1}=d\,.
\eal
The supersymmetry conditions corresponding to the 22 and 33 entries are then automatically satisfied.

As for the central 23 and 32 entries in \eqn{4x4} the difference between the supersymmetry variation 
and the super-covariant derivative of $G$ gives
\bal\label{remainder}
& (\delta\cL - \mathfrak{D}_\tau G)_{23}
=-i\left(c_{(1)\ul1}\bar c_{(1)}^{\ul2} \tilde j_{(1)+}^{\bar A\ul1} + c_{(2)\ul1}\bar c_{(2)}^{\ul2} j_{(2)+}^{\bar A\ul1}\right)\xi_{\bar A\ul2}^-,\\
& (\delta\cL - \mathfrak{D}_\tau G)_{32}
=i\left(c_{(1)\ul2}\bar c_{(1)}^{\ul1} \tilde j_{(1)-}^{\bar A\ul2} + c_{(2)\ul2}\bar c_{(2)}^{\ul1} j_{(2)-}^{\bar A\ul2}\right)\xi_{\bar A\ul1}^+,
\eal
where $\mathfrak{D}_\tau$ is the super-covariant derivative with respect to the super-connection $\cL$.
Imposing the bilinear constraints \eqn{constraints}, we see that $(\delta\cL - \mathfrak{D}_\tau G)_{23}$ 
and $(\delta\cL - \mathfrak{D}_\tau G)_{32}$ are proportional to $j_{(2)+}^{\bar A\ul1}-\tilde j_{(1)+}^{\bar A\ul1}$ and $j_{(2)-}^{\bar A\ul2}-\tilde j_{(1)-}^{\bar A\ul2}$, respectively.
This is similar to the $2\times2$ cases discussed above.

With the constraints \eqn{constraints}, the supersymmetry conditions for the fermionic entries of 
\eqn{4x4} are satisfied, up to a remainder term, which is proportional to 
$(\mu_{(2)})^{\bar A}{}_{\bar B}-(\tilde\mu_{(1)})^{\bar A}{}_{\bar B}$. 
We present the example of the 12 entry.
Consider
\beq
(\mathfrak{D}_\tau G)_{12}=\partial_\tau g_{(1)\ul1}-iA_{(1)1} g_{(1)\ul1} +i g_{(1)\ul1}\cA_{(2)}-i\bar d\epsilon_{\bar A\bar B}q_{(1)}^{\bar A}q_{(2)}^{\bar B} \bar g_{(2)}^{\ul2}.
\eeq
With some manipulation it may be seen that
\bal
\epsilon_{\bar A\bar B}q_{(1)}^{\bar A}q_{(2)}^{\bar B}\bar g_{(2)}^{\ul2}
&=\sqrt{2}c_{(2)\ul2}\left(\nu_{(1)}q_{(1)}^{\bar A}-\frac{1}{2}q_{(1)}^{\bar A}\left(\tilde\nu_{(1)}+\nu_{(2)}\right)\xi_{\bar A\ul1}^+\right.\\
&\quad
\left.-\epsilon_{\bar A\bar B}q_{(1)}^{\bar A}\left((\mu_{(2)})^{\bar B}{}_{\bar C}-(\tilde\mu_{(1)})^{\bar B}{}_{\bar C}\right)\epsilon^{\bar C\bar D}\xi_{\bar D\ul1-}\right).
\eal
This identity, coupled with \eqn{constraints} gives us that
\beq
(\delta\cL - \mathfrak{D}_\tau G)_{12}
=-i\bar d\epsilon_{\bar A\bar B}q_{(1)}^{\bar A}\left((\mu_{(2)})^{\bar B}{}_{\bar C}
-(\tilde\mu_{(1)})^{\bar B}{}_{\bar C}\right)\epsilon^{\bar C\bar D}\xi_{\bar D\ul1-}.
\label{Deltamu}
\eeq
Indeed all fermionic entries have such a remainder term, proportional to 
$(\mu_{(2)})^{\bar B}{}_{\bar C}-(\tilde\mu_{(1)})^{\bar B}{}_{\bar C}$.
We discuss these remainder terms in the next section.

We have also tried to construct Wilson loops with $3\times3$ and $6\times6$ block structure, but 
couldn't find any useful ones. We expect then that the Wilson loops will be given by this connection, 
or that made of a pair of the previous $2\times2$ connections, or a linear combination thereof. In 
any case the Wilson loops will be classified by representations of $U(N_1+N_3|2N_2)$. 

The story for longer segments of the quiver with vanishing CS levels should be similar. The 
even entries will be along the diagonal or in the bifundamental of $U(N_I)\times U(N_{I+2})$ and 
made of a bilinear of the scalars. In the 4-node case the resulting connection would be 
a $U(N_1+2N_3|2N_2+N_4)$ matrix, and the generalization to longer quivers is obvious.

\subsubsection{On $\Delta J$ and $\Delta\mu$}
\label{sec:J}

In the alternating level case we found a pair of superconnections whose variation is a total 
derivative. In the case with a single $k=0$ node we found five possible superconnections but in all 
cases we found an extra term in the variation proportional to $\Delta J$ \eqn{delj}. One can 
add more hypermultiplets and more nodes with vanishing CS levels and the same 
construction leads to superconnections whose variation includes terms proportional to 
$\Delta J$ terms on the different nodes with $k=0$. For the construction with the 
$4\times4$ connection we also found the remainder $\Delta\mu$ term \eqn{Deltamu}

In the case of alternating levels we found a superconnection whose variation doesn't vanish, 
but like in ABJM theory, it is a total derivative. So the variation of the Wilson loop constructed out 
of the superconnection does vanish. In this case, the variation of the Wilson loop built out of any 
of the four superconnections will not vanish, but will rather give the insertion of the integral 
of $\Delta J$ into the Wilson loop.

Still, we expect there to be a BPS Wilson loop associated with this segment of the quiver. 
Indeed, examining the action \cite{im_aux} for the vector multiplets with $k=0$ one notices that 
the gaugino $\lambda$ appears only linearly and multiplying $\Delta J$. Thus the variation of 
the Wilson loop is an insertion of $\frac{\delta}{\delta\lambda}$ into the Wilson loop and the 
path integral over $\lambda$ of any observable will be a total Gra{\ss}mann derivative and 
will therefore vanish. We conclude that the variation of the Wilson loop is zero in the weak 
sense --- all correlation functions with it vanish.

As for the $\Delta\mu$ piece \eqn{Deltamu}. Again the action of the vector multiplet includes 
this combination multiplied by an auxiliary field. Integrating it out identifies 
$(\mu_{(2)})^{\bar B}{}_{\bar C}=(\tilde\mu_{(1)})^{\bar B}{}_{\bar C}$, so this term does not 
obstruct the supersymmetry analysis, and the $4\times4$ loops are thus also supersymmetric 
in the weak sense.

\subsection{Vanishing CS levels: Repeated twisted hypermultiplets}
\label{sec:-k0k}

One last simple case to consider is a segment of the quiver of the form
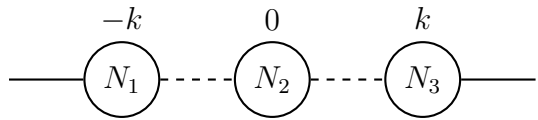
\begin{center}
\begin{tikzpicture}[every node/.style={circle,draw},thick]
\node(NL) at (0,0){$N_1$};
\node(NM) at (2,0){$N_2$};
\node(NR) at (4,0){$N_3$};
\draw[solid](-1.5,0)--(NL);
\draw[dashed](NL)--(NM);
\draw[dashed](NM)--(NR);
\draw[solid](NR)--(5.5,0);
\begin{scope}[nodes = {draw = none,above = 12pt}]
\node at (NR) {$k$};
\node at (NM) {0};
\end{scope}
\begin{scope}[nodes = {draw = none,above = 8pt}]
\node at (NL) {$-k$};
\end{scope}
\end{tikzpicture}
\end{center}

The nodes on the left and on the right will couple to extra nodes through the 
hypermultiplets (solid lines) and form superconnections with them (or if 
we remove the hypermultiplets and consider a linear quiver, will have $1/2$ 
BPS connections like in \eqn{1/4=1/2}). In either case we can construct a 
$1/2$ BPS Wilson loop coupling to the central node alone
\beq
\cL_{(2)}=A_{(2)1}+i(\varphi_{(2)})^{\ul1}{}_{\ul1}
-\frac{i}{k}\left((\mu_{(2)})^{\ul1}{}_{\ul1}+(\tilde\mu_{(1)})^{\ul1}{}_{\ul1}\right).
\eeq

%%%%%%%%%%%%%

\section{Circular Wilson loops}
\label{sec:circ}

For every straight BPS Wilson loop we expect to also find a circular one which will have 
a finite expectation value calculable using the localization matrix model. We study those here.
The circle is given by
\beq
x^1=\cos\tau\,,
\qquad
x^2=\sin\tau\,,
\qquad
x^3=0\,.
\eeq
Whereas the straight line preserved half of the Poincar\'{e} 
supersymmetries and half of the superconformal ones, 
we expect the circular loop to preserve eight linear combinations of the two. 
Concretely, we look for Wilson loops that preserve supersymmetries where 
the superconformal variation parameters $\eta$ are related to the superpoincar\'e 
ones $\xi$ by
\beq\label{c_ans}
\eta_{\bar A\ul1}^a
=i(\sigma^2)^a{}_b \, \xi_{\bar A\ul 1}^b\,,
\qquad
\eta_{\bar A\ul2}^a
=-i(\sigma^2)^a{}_b \, \xi_{\bar A\ul 2}^b\,.
\eeq
The superconformal transformations of the fields are given in the usual way 
\cite{bandres} by replacing $(\xi_{\bar A\ul B})^a\rightarrow x^\mu (\gamma_
\mu)^a{}_b (\eta_{\bar A\ul B})^b$, except for the variations of the fermions 
which pick up an extra term (see \app{not}).

%%%%%%%%%%%%%

\subsection{Alternating levels}

Let us consider the same segment as for the straight line case in Section~\ref{sec:k-k}
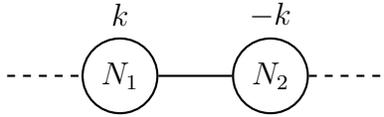
\begin{center}
\begin{tikzpicture}[every node/.style={circle,draw},thick]
\node(NL) at (2,0){$N_1$};
\node(NR) at (4,0){$N_2$};
\draw[dashed](0.5,0)--(NL);
\draw[solid](NL)--(NR);
\draw[dashed](NR)--(5.5,0);
\begin{scope}[nodes = {draw = none,above = 12pt}]
\node at (NL) {$k$};
\end{scope}
\begin{scope}[nodes = {draw = none,above = 8pt}]
\node at (NR) {$-k$};
\end{scope}
\end{tikzpicture}.
\end{center}

As in the straight line case, contribution to the variation of the gauge field from the 
adjoining twisted hypermultiplet may be cancelled by considering the combination 
$A_{(1)1} - \frac{2i}{k}(\tilde\mu_{(0})^{\ul1}{}_{\ul1}$.
Futhermore, we may consider the two combinations
\bal\label{a1cv}
&\delta\left[\dot x^\mu A_{(1)\mu}
-\frac{2i}{k}(\tilde\mu_{(0)})^{\ul1}{}_{\ul1}
-\frac{i}{k}\nu_{(1)}\right]\\
&=\frac{2\sqrt2}{k}\left\{\left[
(1-\sin\tau)\xi_{\bar A\ul1}^+
+\cos\tau\,\xi_{\bar A\ul1}^-\right]
q_{(1)}^{\bar A} \bar\psi_{(1)+}^{\ul1}
%\right.\\&\quad\left.
+\left[\cos\tau\,\xi_{\bar A\ul1}^+
+(1+\sin\tau)\xi_{\bar A\ul1}^-\right]
q_{(1)}^{\bar A} \bar\psi_{(1)-}^{\ul1}
\right.\\&\quad\left.
+\left[(1+\sin\tau)\epsilon^{\bar A\bar B}\epsilon^{\ul1\ul2}\xi^+_{\bar B\ul 2}
-\cos\tau\,\epsilon^{\bar A\bar B}\epsilon^{\ul1\ul2}\xi^-_{\bar B\ul2}\right]
\psi_{(1)\ul1+} \bar q_{(1)}^{\bar A}\right.\\
&\quad\left.+\left[
-\cos\tau\,\epsilon^{\bar A\bar B}\epsilon^{\ul1\ul2}\xi^+_{\bar B\ul2}
+(1-\sin\tau)\epsilon^{\bar A\bar B}\epsilon^{\ul1\ul2}\xi_{\bar B\ul2}\right]
\psi_{(1)\ul1-} \bar q_{(1)}^{\bar A}\right\},
\eal
and
\bal
&\delta\left[\dot x^\mu A_{(1)\mu}
-\frac{2i}{k}(\tilde\mu_{(0)})^{\ul1}{}_{\ul1}+
\frac{i}{k}\nu_{(1)}\right]\\
&=-\frac{2\sqrt2}{k}\left\{\left[
(1+\sin\tau)\xi_{\bar A\ul2}^+
-\cos\tau\,\xi_{\bar A\ul2}^-\right]
q_{(1)}^{\bar A} \bar\psi_{(1)+}^{\ul2}
%\right.\\&\quad\left.
+\left[-\cos\tau\,\xi_{\bar A\ul2}^+
+(1-\sin\tau)\xi_{\bar A\ul2}^-\right]
q_{(1)}^{\bar A} \bar\psi_{(1)-}^{\ul2}\right.\\
&\quad\left.+\left[
(1-\sin\tau)\epsilon^{\bar A\bar B}\epsilon^{\ul2\ul1}\xi^+_{\bar B\ul 1}
+\cos\tau\,\epsilon^{\bar A\bar B}\epsilon^{\ul2\ul1}\xi^-_{\bar B\ul1}\right]
\psi_{(1)\ul2+} \bar q_{(1)}^{\bar A}\right.\\
&\quad\left.+\left[\cos\tau\,\epsilon^{\bar A\bar B}\epsilon^{\ul2\ul1}\xi^+_{\bar B
\ul1}
+(1+\sin\tau)\epsilon^{\bar A\bar B}\epsilon^{\ul2\ul1}\xi^-_{\bar B\ul1}\right]
\psi_{(1)\ul2-} \bar q_{(1)}^{\bar A}\right\}.
\eal
From the straight line case, we expect that these two choices correspond to the $\psi_1$ and 
$\psi_2$-loops respectively.
We assume the form of the superconnections and verify that they are indeed supersymmetric.

\subsubsection{The \texorpdfstring{$\psi_1$}{psi1}-loop}
\label{sec:flc}

We take the superconnection to be of the form
\beq\label{cpsi1}
\cL^{\psi_1}=
\begin{pmatrix}
\dot x^\mu A_{(1)\mu}-\frac{2i}{k}(\tilde\mu_{(0)})^{\ul1}{}_{\ul1}-\frac{i}{k}\nu_{(1)}\quad 
& \bar c_a \psi_{(1)\ul1}^a\\
c^a \bar\psi_{(1)}^{\ul1 a} & \dot x^\mu A_{(2)\mu}-\frac{2i}{k}(\mu_{(2)})^{\ul1}{}_{\ul1}-\frac{i}{k}\tilde\nu_{(1)}
\end{pmatrix},
\eeq
and the supermatrix $G$ to be of the form
\beq
\label{gcirc}
G=
\begin{pmatrix}
0 & g\\
\bar g & 0
\end{pmatrix}.
\eeq

As in \cite{drukker}, we consider the projector
\beq
\cP^+\equiv
\delta^a{}_b+\dot x^\mu (\gamma_\mu)^a{}_b=
\begin{pmatrix}
1-\sin\tau & \cos\tau\\
\cos\tau & 1+\sin\tau
\end{pmatrix},
\eeq
and demand that the fermionic couplings be eigenstates of this projector.
In particular, we choose
\beq
c^a=\left(\cos\tau,\,1+\sin\tau\right)^a c(\tau)\,,
\qquad
\bar c_a=
\begin{pmatrix}
1-\sin\tau\\
\cos\tau
\end{pmatrix}_a
\bar c(\tau),
\eeq
where $c(\tau)$ and $\bar c(\tau)$ are functions of $\tau$.	
The variation is given by
\bal
\delta\left[\bar c_a \psi_{(1)\ul1}^a\right]
=&\,2\sqrt{2}D_\tau q^{\bar A} \bar c_a \xi_{\bar A\ul1}^a
-\frac{2\sqrt{2}}{k}\left[
(\nu_{(1)})^{\bar B}{}_{\bar B} q^{\bar A}_{(1)}
-q^{\bar A}_{(1)} (\tilde\nu_{(1)})^{\bar B}{}_{\bar B}
\right]\bar c_a \xi_{\bar A\ul1}^a\\
&+i\sqrt2\bar c_a (\sigma^2)^a{}_b \xi_{\bar A\ul1}^b q_{(1)}^{\bar A}.
\eal
We wish to write this as a covariant derivative.
We may impose
\beq
\partial_\tau \bar c_a=\frac{i}{2}\bar c_b (\sigma^2)^b{}_a,
\eeq
which tells us that
\beq
\label{barctau}
\bar c(\tau)=\frac{\bar C}{\cos\frac{\tau}{2}-\sin\frac{\tau}{2}},
\eeq
where $\bar C$ is a constant.
With this choice of $\bar c(\tau)$, we have
\beq
\delta\left[\bar c_a (\psi_{(1)\ul1})^a\right]
=\cD_\tau\left(2\sqrt{2}q^{\bar A} \bar c_a \xi_{\bar A\ul1}^a\right),
\eeq
where $\cD_\tau q_{(1)}^{\bar A}\equiv D_\tau q^{\bar A}+\frac{1}{k}
\nu_{(1)} q^{\bar A}-\frac{1}{k}q^{\bar A}\tilde
\nu_{(1)}$.
The supersymmetry conditions demand
\beq
g=2\sqrt{2}q_{(1)}^{\bar A}\bar c_a \xi_{\bar A\ul1}^a.
\eeq 
Similarly, we find
\beq
\delta\left[c^a (\bar\psi_{(1)}^{\ul1})_a\right]
=\cD_\tau \left(-2\sqrt{2}\bar q_{(1)\bar A} c^a \epsilon^{\bar A\bar B}
\epsilon^{\ul1\ul2}\xi_{\bar B\ul2 a}\right),
\eeq
where we have imposed that
\beq
\partial_\tau c^a=-\frac{i}{2}c^b(\sigma^2)^a{}_b,
\eeq
which is solved by
\beq
\label{ctau}
c(\tau)=\frac{C}{\cos\frac{\tau}{2}+\sin\frac{\tau}{2}},
\eeq
where $C$ is a constant.
We thus have
\beq
\bar g=-2\sqrt{2}\bar q_{(1)\bar A}c^a \epsilon^{\bar A\bar B}
\epsilon^{\ul1\ul2}\xi_{\bar B\ul2a}.
\eeq
A simple calculation, as in the straight line case in Section~\ref{sec:k0l1} gives that $C\bar C=-\frac{i}{k}$.

The story for the segment of the quiver with vanishing CS level is very similar, where for example 
the circular analog of the straight line presented in Section~\ref{sec:k0l1} is the same as 
\eqn{cpsi1} with the lower right corner replaced by
$\dot x^\mu A_{(2)\mu}+i(\varphi_{(2)})^{\ul1}{}_{\ul1}-\frac{i}{k}\tilde\nu_{(1)}$, which 
is essentially the same as in \eqn{k0-kpsi11}.
The variation of the super-connection is then
\beq
\delta\cL^{\psi_1}\equiv \mathfrak{D}_{\tau}^{\psi_1} G + \Delta J
\begin{pmatrix}
0 & 0\\
0 & 1
\end{pmatrix},
\eeq
with $G$ as in \eqn{gcirc} and $\Delta J$ is as defined in \eqn{delj}.
The other loops for this case may be similarly obtained in analogy with the straight line case.

%%%%%%%%%%%%%

\subsubsection{The \texorpdfstring{$\psi_2$}{psi2}-loop}
\label{sec:f2c}

To construct a loop coupling to $\psi_{\ul2}$, we take the fermionic couplings to be eigenstates of the projector 
$\cP^-=\delta^a{}_b - \dot x^\mu (\gamma_\mu)^a{}_b$, given explicitly by
\beq
\cP^-=
\begin{pmatrix}
1+\sin\tau & -\cos\tau\\
-\cos\tau & 1-\sin\tau
\end{pmatrix}.
\eeq
We take the fermionic couplings to be
\beq
c^a=c(\tau)
\begin{pmatrix}
-\cos\tau, & 1-\sin\tau
\end{pmatrix}^a,
\qquad
\bar c_a
=\bar c(\tau)
\begin{pmatrix}
1+\sin\tau\\
-\cos\tau
\end{pmatrix}_a,
\eeq
with the superconnection given by
\beq\label{cpsi2}
\cL^{\psi_2}
=
\begin{pmatrix}
\dot x^\mu A_{(1)\mu}-\frac{2i}{k}(\tilde\mu_{(0)})^{\ul1}{}_{\ul1}
+\frac{i}{k}\nu_{(1)}\quad 
& \bar c_a \psi_{(1)\ul2}^-
\\
c^a \bar\psi_{(1)-}^{\ul2} 
& \dot x^\mu A_{(2)\mu}-\frac{2i}{k}(\mu_{(2)})^{\ul1}{}_{\ul1}
+\frac{i}{k}\tilde\nu_{(1)}
\end{pmatrix}.
\eeq
Proceeding as in the previous section, studying the variations of the fermions 
gives us
\bal
g&=-2\sqrt{2}q^{\bar A}\bar c_a (\xi_{\bar A\ul2})^a\,,
&\qquad
\bar c(\tau)&=\frac{\bar C}{\cos\frac{\tau}{2}+\sin\frac{\tau}{2}}\,,
\\
\bar g&=2\sqrt{2}\,\bar q_{\bar A} c^a (\xi^{\bar A\ul2})_a\,,
&\qquad
c(\tau)&=\frac{C}{\cos\frac{\tau}{2}-\sin\frac{\tau}{2}}\,,
\eal 
as well as compatibility with the conjectured bosonic connection of the two nodes.
An examination of the variation of the bosonic connections gives $C\bar C=\frac{i}
{k}$.

\subsubsection{The boundary conditions}

We have shown that the supersymmetric variation of $\cL$ may be written as a 
super-covariant derivative. In order for the circle, which is compact, to be invariant under 
supersymmetry we must ensure that there are no boundary terms when integrating this 
total derivative term.
We follow the analysis in \cite{cardinali}.

We define $\mathfrak{W}$ to be the untraced Wilson loop and analyze its transformation by the 
total derivative viewed as a super gauge transformation.
Consider the matrix $G$ (for either the $\psi_1$ and $\psi_2$-loops) found above, then we find
\beq
G(2\pi)=-G(0)\,.
\eeq
This can be written as
\beq
G(2\pi)=\mathcal{T}G(0)\mathcal{T}^{-1}\,,
\qquad
\text{with}
\qquad
\mathcal{T}
=
\begin{pmatrix}
i & 0\\
0 & -i
\end{pmatrix}\,.
\eeq
This implies that the finite gauge transformation $U=\exp(iG)$ also satisfies
\beq
U(2\pi)=\mathcal{T}U(0)\mathcal{T}^{-1}\,.
\eeq

Now we consider the transformation of $\mathfrak{W}\mathcal{T}$. It is invariant under the
supersymmetry/super\-gauge transformations. 
\beq
\sTr(\mathfrak{W}\mathcal{T})
\rightarrow
\sTr(U^{-1}(0)\mathfrak{W} U(2\pi)\mathcal{T})
=\sTr(U^{-1}(0)\mathfrak{W} \mathcal{T} U(0))
=\sTr(\mathfrak{W} \mathcal{T})\,.
\eeq
We thus should consider the supertrace of $\mathfrak{W}\mathcal{T}$ rather than of $\mathfrak{W}$.
Equvalently, examining that form of $\mathcal{T}$, we see that 
the trace of $\mathfrak{W}$ (rather than the supertrace) is a supersymmetric operator. 
This is similar to the result for ABJM in \cite{drukker}.

%%%%%%%%%%%%%%

\section{Localization and the matrix model}
\label{sec:loc}

We would now like to show that we can calculate the expectation value of the $1/2$ 
BPS Wilson loops using localization. The idea, as in \cite{drukker}, is to show that the 
$1/2$ BPS loops are cohomologically equivalent, under the localization supercharge, 
to certain combinations of $1/4$ BPS loops for which the localization calculation has 
already been done. This reduces the calculation of the Wilson loop to that of an observable 
in a matrix model, as in 
\cite{putrov,marino,klemm,moriyama}. We shall 
leave the solution of the matrix model to future work.

\subsection{The straight line}

We now demonstrate, following closely \cite{drukker}, that both the $\psi_1$-loops and $\psi_2$-loops 
are in the same cohomology class --- that of a certain $1/4$ BPS straight loop. We should note that 
this is a classical calculation and we expect it to be modified by quantum corrections and in fact only 
a single linear combination of these loops will be exactly equivalent to the $1/4$ BPS loop. We discuss 
the lifting of the degeneracy in Section~\ref{sec:disc}.

The appropriate $1/4$ BPS operator has a purely bosonic connection from two adjacent nodes
\beq
\label{1/4}
\cL^{1/4}=
\begin{pmatrix}
A_{(1)1} - \frac{2i}{k}\left((\mu_{(0)})^{\ul1}{}_{\ul1}+(\mu_{(1)})^{\bar1}{}_{\bar1}\right) & 0\\
0 & A_{(2)1} - \frac{2i}{k}\left((\tilde\mu_{(1)})^{\bar1}{}_{\bar1}+(\mu_{(2)})^{\ul1}{}_{\ul1}\right)
\end{pmatrix}.
\eeq
We note that this shares its four supersymmetries with our $1/2$ BPS loops. 
As our localizing supercharge, we consider the linear combination
\beq
Q\equiv Q^{\bar1\ul2}_- + Q^{\bar2\ul1}_+.
\eeq

\subsubsection{The $\psi_1$-loop}

The difference between the $\psi_1$-loop \eqn{lpsi1} found in Section~\ref{sec:a1} and the $1/4$ BPS loop is given by
\beq
\cW^{\psi_1}-\cW^{1/4}
=\tr\cP\left[e^{-i\int d\tau\,\cL^{\psi_1}} - e^{-i\int d\tau\,\cL^{1/4}}\right].
\eeq
We wish to demonstrate that this is a $Q$-exact quantity by finding a $V$ such that $\cW^{\psi_1}-\cW^{1/4}=QV$.
Following \cite{drukker}, we take 
\beq
\label{V1}
V^{\psi_1}=-i\tr\cP\left[\int_{-\infty}^\infty d\tau\,e^{-i\int_{-\infty}^\tau d\tau_1\, \cL^{1/4}} \Lambda^{\psi_1}(\tau) e^{-i\int_\tau^\infty d\tau_2\,\cL^{\psi_1}} \right],
\eeq
where $\Lambda^{\psi_1}$ is to be determined by the requirement $Q\Lambda^{\psi_1}=\cL_F^{\psi_1}$ and $\cL_F^{\psi_1}$ is the connection with fermionic entries equal to that of $\cL^{\psi_1}$ and zeros everywhere else.
We find
\beq
\label{Lambda1}
\Lambda^{\psi_1}=\frac{i}{\sqrt{2}}
\begin{pmatrix}
0 & \bar c\,q_{(1)}^{\bar2}\\
-c\,\bar q_{(1)\bar2} & 0
\end{pmatrix},
\qquad
(\Lambda^{\psi_1})^2
=\frac{1}{2}\tilde\cL^{\psi_1}_B\equiv\frac{1}{2}\left(\cL^{\psi_1}_B-\cL^{1/4}_B\right),
\eeq
where the subscript `B' indicates the bosonic part of and $\tilde\cL^{\psi_1}\equiv\cL^{\psi_1}-\cL^{1/4}$.

Acting with $Q$ on $V^{\psi_1}$ gives
\beq
Q V^{\psi_1}=-i\tr\cP\left[\int_{-\infty}^\infty d\tau\,e^{-i\int_{-\infty}^\tau d\tau_1\, 
\cL^{1/4}} \left(\cL_F^{\psi_1}(\tau) + \Lambda^{\psi_1}Q\right) 
e^{-i\int_\tau^\infty d\tau_2\,\cL^{\psi_1}} \right].
\eeq
We pick up a boundary term from acting with the supercharges on the exponent.
Noting that
\beq
\label{Qpsi1}
Q\psi_{(1)\ul1}^+=\sqrt{2}\cD_\tau q_{(1)}^{\bar2}\,,
\qquad 
Q\bar\psi_{(1)+}^{\ul1}=-\sqrt{2}\cD_\tau \bar q_{(1)\bar2}\,,
\eeq
we find
\bal
Q V^{\psi_1}&=-i\tr\cP\left[\int_{-\infty}^\infty d\tau\,e^{-i\int_{-\infty}^\tau d\tau_1\, \cL^{1/4}} \left(\cL_F^{\psi_1}(\tau) + \tilde\cL^{\psi_1}_B\right) e^{-i\int_\tau^\infty d\tau_2\,\cL^{\psi_1}} \right],\\
&=-i\tr\cP\left[\int_{-\infty}^\infty d\tau\,e^{-i\int_{-\infty}^\tau d\tau_1\, \cL^{1/4}} \tilde\cL^{\psi_1}(\tau) e^{-i\int_\tau^\infty d\tau_2\,\cL^{\psi_1}} \right].
\eal

As in \cite{drukker}, upon Taylor expansion, we see that this is equal to the expansion of
\beq
\cW^{\psi_1}-\cW^{1/4}
=\tr\cP\left[e^{-i\int_{-\infty}^\infty \cL^{1/4}} \sum_{p=1}^\infty 
(-i)^p \int d\tau_1 \cdots d\tau_p\, \tilde\cL^{\psi_1}(\tau_1)\cdots\tilde\cL^{\psi_1}(\tau_p)\right],
\eeq
demonstrating the cohomological equivalence of the $\psi_1$-loop and $1/4$ BPS loop.

\subsubsection{The $\psi_2$-loop}

The $\psi_2$-loop \eqn{lpsi2} of Section~\ref{sec:p2} is also cohomologically equivalent to the $1/4$ BPS loop.
The result follows almost identically to that of the $\psi_1$-loop, with
\beq
V^{\psi_2}=-i\tr\cP\left[\int_{-\infty}^\infty d\tau\,
e^{-i\int_{-\infty}^\tau d\tau_1\, \cL^{1/4}} \Lambda^{\psi_2}(\tau) 
e^{-i\int_\tau^\infty d\tau_2\,\cL^{\psi_1}} \right],
\eeq
where
\beq
\label{Lambda2}
\Lambda^{\psi_2}=-\frac{i}{\sqrt{2}}
\begin{pmatrix}
0 & \bar c\,q_{(1)}^{\bar1}\\
c\,\bar q_{(1)\bar1} & 0
\end{pmatrix},
\qquad
(\Lambda^{\psi_1})^2
=\frac{1}{2}\tilde\cL^{\psi_2}_B
\equiv\frac{1}{2}\left(\cL^{\psi_2}_B-\cL^{1/4}_B\right).
\eeq
Noting also that
\beq
\label{Qpsi2}
Q\psi_{(1)\ul2}^-=-\sqrt{2}\cD_\tau q_{(1)}^{\bar1}\,,
\qquad 
Q\bar\psi_{(1)-}^{\ul2}=-\sqrt{2}\cD_\tau \bar q_{(1)\bar1}\,,
\eeq
it is straightforward to see that both loops are cohomologically equivalent to each other.

\subsubsection{Vanishing CS levels}

In the case of the quivers with vanishing CS levels discussed in Section~\ref{sec:k0-k}
We have not demonstrated the cohomological equivalence to $1/4$ BPS loops for 
the loops based on $2\times2$ blocks. Below we do it for the $4\times4$ superconnection 
\eqn{4x4}. For the $1/2$ BPS loop in Section~\ref{sec:-k0k} it is in fact identical to the $1/4$ BPS loop.

We found the $U(N_1+N_3|2N_2)$ valued superconnection \eqn{4x4} in Section~\ref{sec:4x4} 
We now apply the arguments of the previous section to this loop. 
It is useful to rearrange the superconnection to make manifest its $U(N_1+N_3|2N_2)$ structure:
\beq
\cL=
\begin{pmatrix}
A_{(1)1} & \bar d\epsilon_{\bar A\bar B} q_{(1)}^{\bar A}q_{(2)}^{\bar B} 
& \bar c_{(1)}^{\ul1}\psi_{(1)\ul1}^+ 
& \bar c_{(1)}^{\ul2}\psi_{(1)\ul2}^-\\
d\epsilon^{\bar A\bar B}\bar q_{(2)\bar A}q_{(1)\bar B}
& A_{(3)1}
& c_{(2)\ul2}\bar\psi_{(2)-}^{\ul2} & c_{(2)\ul1}\bar\psi_{(2)+}^{\ul1}\\
c_{(1)\ul1}\bar\psi_{(1)+}^{\ul1} & \bar c_{(2)}^{\ul2} \psi_{(2)\ul2}^-
& \cA_{(2)}
& 0\\
c_{(1)\ul2}\bar\psi_{(1)-}^{\ul2} 
& \bar c_{(2)}^{\ul1}\psi_{(2)\ul1}^+ & 0
& \cA_{(2)}^\prime
\end{pmatrix}.
\eeq
As in \eqn{V1} we define
\beq
V^{4\times4}=-i\tr\cP\left[\int_{-\infty}^\infty d\tau\,
e^{-i\int_{-\infty}^\tau d\tau_1\, \cL^{1/4}} 
\Lambda^{4\times4}(\tau) e^{-i\int_\tau^\infty d\tau_2\,\cL^{\psi_1}} \right],
\eeq
where $\cL^{1/4}$ is the combination of $1/4$ BPS connections
\begin{align}
\cL^{1/4}&=
\tx{diag}\left\{A_{(1)1}-\frac{2i}{k}(\mu_{(1)})^{\bar1}{}_{\bar1},
A_{(3)1}-\frac{2i}{k}(\tilde\mu_{(2)})^{\bar1}{}_{\bar1},\right.\\
&\quad\left.A_{(2)1}+i(\varphi_{(2)})^{\ul1}{}_{\ul1}-\frac{i}{k}\left((\tilde\mu_{(1)})^{\bar1}{}_{\bar1}+
(\mu_{(2)})^{\bar1}{}_{\bar1}\right),\,
A_{(2)1}+i(\varphi_{(2)})^{\ul1}{}_{\ul1}-\frac{i}{k}\left((\tilde\mu_{(1)})^{\bar1}{}_{\bar1}+(\mu_{(2)})^{\bar1}{}_{\bar1}\right)\right\}.
\nonumber
\end{align}
We determine $\Lambda^{4\times4}$ by the requirement that $Q\Lambda^{4\times4}=\cL_F$, with 
$\cL_F$ the fermionic part of $\cL$.

Given the same supercharge and many of the same fermionic entries as in the examples above, 
we may read build $\Lambda^{4\times4}$ from \eqn{Lambda1}, \eqn{Lambda2}
\beq
\Lambda^{4\times4}=-\frac{i}{\sqrt{2}}
\begin{pmatrix}
0 & 0 & -\bar c_{(1)}^{\ul1} q_{(1)}^{\bar2} & \bar c_{(1)}^{\ul2}q_{(1)}^{\bar2}\\
0 & 0 & c_{(2)\ul2}\bar q_{(2)\bar1} & c_{(2)\ul1} \bar q_{(2)\bar2}\\
c_{(1)\ul1} \bar q_{(1)\bar2} & \bar c_{(1)}^{\ul2} q_{(2)}^{\bar1} & 0 & 0\\
c_{(1)\ul2}\bar q_{(1)\bar1} & -\bar c_{(1)}^{\ul1} q_{(2)}^{\bar2} & 0 & 0
\end{pmatrix}.
\eeq
A short computation ( making use of the constraints \eqn{constraints}) shows that
\beq
(\Lambda^{4\times4})^2=\frac{1}{2}\tilde\cL_B -\frac{1}{2}
\begin{pmatrix}
0 & 0 & 0 & 0\\
0 & 0 & 0 & 0\\
0 & 0 & 0 
& c_{(1)\ul1}\bar c_{(1)}^{\ul2}\left((\mu_{(2)})^{\bar1}{}_{\bar2} 
- (\tilde\mu_{(1)})^{\bar1}{}_{\bar2}\right)\\
0 & 0 & c_{(1)\ul2}\bar c_{(1)}^{\ul1}\left((\mu_{(2)})^{\bar2}{}_{\bar 1}
-(\tilde\mu_{(1)})^{\bar2}{}_{\bar 1}\right) & 0
\end{pmatrix}.
\eeq
As before, $\tilde\cL_B$ is the bosonic part of the difference $\cL-\cL^{1/4}$.
We note that both of the non-zero entries are proportional to $(\mu_{(2)})^{\bar A}{}_{\bar B}
-(\tilde\mu_{(1)})^{\bar A}{}_{\bar B}$, for fixed $\bar A$, $\bar B$.
For notational convenience, we define $(\Lambda^{4\times4})^2\equiv \frac{1}{2}\tilde\cL_B+\Delta\mu$.

Continuing as before, using the action of the supercharges 
on the fermions given in \eqn{Qpsi1} and \eqn{Qpsi2}, we see that
\beq
Q\cL_F=-2i\mathfrak{D}_\tau\Lambda,
\eeq
where $\mathfrak{D}_\tau$ is the supercovariant derivative with respect to $\cL$.

Acting with the supercharge on $V$ thus gives
\bal
Q V&=-i\tr\cP\left[\int_{-\infty}^\infty d\tau\,e^{-i\int_{-\infty}^\tau d\tau_1\, 
\cL^{1/4}} \left(\cL_F(\tau) + \Lambda Q\right) 
e^{-i\int_\tau^\infty d\tau_2\,\cL} \right]\\
&=-i\tr\cP\left[\int_{-\infty}^\infty d\tau\,e^{-i\int_{-\infty}^\tau d\tau_1\, \cL^{1/4}} \left(\cL_F(\tau) + \tilde\cL_B + \Delta\mu\right) e^{-i\int_\tau^\infty d\tau_2\,\cL^{\psi_1}} \right]\\
&=-i\tr\cP\left[\int_{-\infty}^\infty d\tau\,e^{-i\int_{-\infty}^\tau d\tau_1\, \cL^{1/4}} \left(\tilde\cL(\tau) + \Delta\mu\right) e^{-i\int_\tau^\infty d\tau_2\,\cL} \right].
\eal
As in the previous cases, the first term in brackets matches the expansion of 
$\cW-\cW^{1/4}$ exactly, where $\cW$ is the loop defined with superconnection $\cL$.
The second term is however new.

In Section~\ref{sec:J}, we argued that the variation of the loop \eqn{4x4} 
vanishes inside correlation functions. Indeed there too we found terms proportional to 
$\mu_{(2)}-\tilde\mu_{(1)}$, which vanish once the auxiliary field is integrated out. 
This is enough to conclude that the expectation value of $\cW$ and $\cW^{1/4}$ are 
equal.

%%%%%%%%%%%%%

\subsection{The circle}

We now consider the circle. 
This will be cohomologically equivalent to the circular analogs of the $1/4$ 
BPS loops above, which have a finite expectation value, which can be calculated by 
localization \cite{kapustin}.
We take $Q$ to be the same supercharge used for localization
\beq
Q=(Q^{\bar1\ul2}_- - S^{\bar1\ul2}_+) + (Q^{\bar2\ul1}_- + S^{\bar2\ul1}_+).
\eeq

\subsubsection{The $\psi_1$-loop}

We begin by considering the difference $\cW^{\psi_1}-\cW^{1/4}$ between the $\psi_1$-loop \eqn{cpsi1} found in Section~\ref{sec:flc} and the $1/4$ BPS circle, similar to \eqn{1/4}. 
We again look for $V^{\psi_1}$ as in \eqn{V1}, this time perturbatively in a power series. 
It will be useful to consider the insertion
\beq
\Lambda^{\psi_1}=\frac{i}{\sqrt{2}}
\begin{pmatrix}
0 & \bar c(\tau) q_{(1)}^{\bar2}\\
-c(\tau) \bar q_{(1)\bar2} & 0
\end{pmatrix},
\eeq
with $c(\tau)$ and $\bar c(\tau)$ as in \eqn{ctau}, \eqn{barctau}). 
It satisfies
\beq\label{con_1p}
Q\Lambda^{\psi_2}=\cL_F^{\psi_2}\,,
\qquad
4\cos\tau (\Lambda^{\psi_1})^2
=\tilde\cL^{\psi_1}_B\,.
\eeq
Furthermore, the action of the supercharges on 
the fermionic components of the $\psi_1$-loop is given by
\beq\label{con_2p}
Q\cL_F^{\psi_1}=2\sqrt{2}\cD_\tau\left[\cos\tau
\begin{pmatrix}
0 & \bar c(\tau) q^{\bar2}_{(1)}\\
-c(\tau) \bar q_{(1)\bar2} & 0
\end{pmatrix}
\right]
=-4i\cD_\tau\left[\cos\tau\,\Lambda^{\psi_1}\right].
\eeq

Order-by-order, we define
\bal
V^{\psi_1}_1&=-i\tr\cP\left\{e^{-i\int_0^{2\pi}d\tau_1\,\cL^{1/4}(\tau_1)}
\int_0^{2\pi}d\tau_2\,\Lambda^{\psi_1}\right\},
\\
V^{\psi_1}_2&=-\frac{1}{2}\tr\cP\left\{e^{-i\int_0^{2\pi}d\tau_1\,\cL^{1/4}(\tau_1)}
\int_0^{2\pi}d\tau_2\int_{\tau_2}^{2\pi}d\tau_3\,
\left(\Lambda^{\psi_1}(\tau_2)\cL^{\psi_1}_F(\tau_3) - \cL^{\psi_1}_F(\tau_2)
\Lambda^{\psi_1}(\tau_3)\right)\right\}
\\
V^{\psi_1}_3&=\tr\cP\left\{e^{-i\int_0^{2\pi}d\tau_1\,\cL^{1/4}(\tau_1)}
\left(-\int_0^{2\pi}d\tau_2\int_{\tau_2}^{2\pi}d\tau_3
\left(\tilde\cL^{\psi_1}_B(\tau_2)\Lambda^{\psi_1}(\tau_3) + 
\Lambda^{\psi_1}(\tau_2)\tilde\cL^{\psi_1}_B(\tau_3)\right)\right.\right.
\\
&\quad\left.\left.+i\int_{\tau_2<\tau_3<\tau_4}d\tau_2\,d\tau_3\,d\tau_4\,\left(\Lambda^{\psi_1}(\tau_2)\cL_F^{\psi_1}(\tau_3)\cL_F^{\psi_1}(\tau_4) + \cL_F^{\psi_1}(\tau_2)\Lambda^{\psi_1}(\tau_3)\cL_F^{\psi_1}(\tau_4)\right.\right.\right.\\
&\quad\left.\left.\left. + \cL_F^{\psi_1}(\tau_2)
\cL_F^{\psi_1}(\tau_3)\Lambda^{\psi_1}(\tau_4)\right)\right)\right\}.
\eal
The action of the supercharges on these operators is given by
\bal
QV^{\psi_1}_1&=-i\tr\cP\left\{e^{-i\int_0^{2\pi}d\tau_1\,\cL^{1/4}(\tau_1)}
\int_0^{2\pi}d\tau_2\,\cL^{\psi_1}_F\right\},
\\
\label{qv2} 
QV^{\psi_1}_2&=-\tr\cP\left\{e^{-i\int_0^{2\pi}d\tau_1\cL^{1/4}(\tau_1)}\int_0^{2\pi}d\tau_2\left(\int_{\tau_2}^{2\pi}d\tau_3\,\cL_F^{\psi_1}(\tau_2)\cL^{\psi_1}_F(\tau_3)+i\tilde\cL^{\psi_1}_B(\tau_2)\right)\right\},
\\
QV^{\psi_1}_3&=\tr\cP\left\{e^{-i\int_0^{2\pi}d\tau_1\,\cL^{1/4}(\tau_1)}\left(-\int_{\tau_2<\tau_3}d\tau_2d\tau_3\,\left(
\tilde\cL^{\psi_1}_B(\tau_2)\cL_F^{\psi_1}(\tau_3) 
+\cL_F^{\psi_1}(\tau_2)\tilde\cL^{\psi_1}_B(\tau_3)\right)\right.\right.\\
&\quad\left.\left.+i\int_{\tau_2<\tau_3<\tau_4}d\tau_2\,d\tau_3\,d\tau_4\,\cL^{\psi_1}_F(\tau_2)\cL_F^{\psi_1}(\tau_3)\cL_F^{\psi_1}(\tau_4)\right)\right\}.
\eal
These indeed match the terms in the expansion of $\cW^{\psi_1}-\cW^{1/4}$, 
demonstrating that to third order in the expansion the $\psi_1$-loop is cohomologically 
equivalent to the $1/4$ BPS loop.
However, as in the case of ABJM, we expect this relationship to hold to all orders, and 
the corresponding localization calculation to reduce to the known matrix model calculation 
of the $1/4$ BPS loops \cite{kapustin}.

%%%%%%%%%

\subsubsection{The $\psi_2$-loop}

The calculation for the $\psi_2$-loop, \eqn{cpsi2} found in Section~\ref{sec:f2c}, proceeds very similarly. Defining
\beq
\Lambda^{\psi_2}=\frac{i}{\sqrt{2}}
\begin{pmatrix}
0 & \bar c(\tau) q_{(1)}^{\bar1}\\
-c(\tau) \bar q_{(1)\bar1} & 0
\end{pmatrix},
\eeq
we see that it satisfies
\beq\label{con_1}
Q\Lambda^{\psi_2}=\cL_F^{\psi_2}\,,
\qquad
4\cos\tau (\Lambda^{\psi_2})^2
=\tilde\cL^{\psi_2}_B\,,
\eeq
and
\beq\label{con_2}
Q\cL_F^{\psi_2}=2\sqrt{2}\cD_\tau\left[\cos\tau
\begin{pmatrix}
0 & \bar c(\tau) q^{\bar1}_{(1)}\\
-c(\tau) \bar q_{(1)\bar1} & 0
\end{pmatrix}
\right]
=-4i\cD_\tau\left[\cos\tau\,\Lambda^{\psi_2}\right].
\eeq
Equations \eqn{con_1} and \eqn{con_2} are the corresponding identities for the $\psi_2$-loop as \eqn{con_1p} and \eqn{con_2p} for the $\psi_1$-loop.
As these are identical, it immediately follows that the action of the supercharge on
\begin{align}
V^{\psi_2}_1&=-i\tr\cP\left\{e^{-i\int_0^{2\pi}d\tau_1\,\cL^{1/4}(\tau_1)}
\int_0^{2\pi}d\tau_2\,\Lambda^{\psi_1}\right\},
\nonumber\\
V^{\psi_2}_2&=-\frac{1}{2}\tr\cP\left\{e^{-i\int_0^{2\pi}d\tau_1\,\cL^{1/4}(\tau_1)}
\int_0^{2\pi}d\tau_2\int_{\tau_2}^{2\pi}d\tau_3\,
\left(\Lambda^{\psi_1}(\tau_2)\cL^{\psi_1}_F(\tau_3) - \cL^{\psi_1}_F(\tau_2)
\Lambda^{\psi_1}(\tau_3)\right)\right\}
\nonumber\\
V^{\psi_2}_3&=\tr\cP\left\{e^{-i\int_0^{2\pi}d\tau_1\,\cL^{1/4}(\tau_1)}
\left(-\int_0^{2\pi}d\tau_2\int_{\tau_2}^{2\pi}d\tau_3
\left(\tilde\cL^{\psi_1}_B(\tau_2)\Lambda^{\psi_1}(\tau_3) + 
\Lambda^{\psi_1}(\tau_2)\tilde\cL^{\psi_1}_B(\tau_3)\right)\right.\right.
\nonumber\\
&\quad\left.\left.+i\int_{\tau_2<\tau_3<\tau_4}d\tau_2\,d\tau_3\,d\tau_4\,\left(\Lambda^{\psi_1}(\tau_2)\cL_F^{\psi_1}(\tau_3)\cL_F^{\psi_1}(\tau_4) + \cL_F^{\psi_1}(\tau_2)\Lambda^{\psi_1}(\tau_3)\cL_F^{\psi_1}(\tau_4)\right.\right.\right.
\nonumber\\
&\quad\left.\left.\left. + \cL_F^{\psi_1}(\tau_2)\cL_F^{\psi_1}(\tau_3)\Lambda^{\psi_1}(\tau_4)\right)\right)\right\},
\end{align}
will exactly match the expansion of $\cW^{\psi_2}-\cW^{1/4}$, thus demonstrating 
the equivalence of the two loops with respect to the localizing supercharge.

%%%%%%%%%%%%%%%%%

\section{Holographic description}
\label{sec:M-theory}

The goal of this section is to find the M2-brane configurations which preserve 
half of the supersymmetries of the vacuum and are the holographic duals to the 
gauge theory operators discussed above. 

We limit our attention to the case of circular quivers whose corresponding space-time metric  is 
$AdS_4\times M_{p,q,k}$ \cite{imamura}, where the compact manifold is an orbifold of the 7-sphere 
\beq
M_{p,q,k}=\left(S^7/(\bZ_p\oplus \bZ_q)\right)/\bZ_k\,,
\eeq
with radius $R^6=2^5\pi^2 N k p q $ (in units of $\ell_s=1$).  
The gauge group rank $N$ corresponds here to the number of M2-branes located at the orbifold point.

For the $AdS_4$ factor, we consider a metric with an $AdS_2$ foliation
\be
ds^2_{AdS_4}=du^2+\cosh^2u\, ds^2_{AdS_2}+\sinh^2 u\, d\phi^2\,,
\ee
where for a time-like (straight) Wilson loop we take a Lorentzian $AdS_2$
\be
ds^2_{AdS_2}=d\rho^2-\cosh^2\rho\, dt^2\,,
\ee
and for the space-like circle we consider the Euclidean metric
\be
ds^2_{AdS_2}=d\rho^2+\sinh^2\rho\, d\psi^2\,.
\ee
The M2-brane world-volume will be along these $AdS_2$ coordinates in $AdS_4$, while 
on the internal space it will wrap an $S^1$ ({\it i.e.}, the M-theory circle) and sit at a fixed 
point in the remaining coordinates, which is determined by supersymmetry, as we shall see. 

The orbifold action giving rise to the space $M_{p,q,k}$ is most easily described 
in terms of complex embedding coordinates $z_i$ ($i=1,\ldots,4$) subject to the constraint 
\be
\sum_{i=1}^4 |z_i|^2=1\,,
\label{sphere-constraint}
\ee
and to the identifications \cite{imamura}
\be
(z_1,z_2,z_3,z_4)\sim
(\omega_{kp}^m z_1,\omega_{kp}^m z_2,\omega_{kq}^{-m}z_3,\omega_{kq}^{-m}z_4)\,,\qquad
\omega_r=e^{2\pi i/r}\,,\qquad m\in \mathbb{Z}\,.
\label{orb_full}
\ee
This last line alone without the constraint (\ref{sphere-constraint}) would define the manifold
\be
{\cal M}_{p,q,k}=\left(\mathbb{C}^2/\mathbb{Z}_p\otimes \mathbb{C}^2/ \mathbb{Z}_q\right)/\mathbb{Z}_k\,.
\ee
The $SU(2)_{A}$ subgroup of the $R$-symmetry acts on $\mathbb{C}^2/\mathbb{Z}_{p}$ 
and $SU(2)_B$ acts on $\bC^2/\bZ_q$.

The supersymmetries preserved by the Wilson loops \eqn{ta}
are doublets of the $R$-symmetry subgroup $SU(2)_B$, which is left unbroken. 
We therefore expect that the M2-brane embedding will be at a fixed point of the 
$SU(2)_B$ action. This is the tip of the $\bC/\bZ_q$ cone in $\cM_{p,q,k}$.
A similar story applies of course 
if we were to consider loops invariant under $SU(2)_A$ with the two cones exchanged.

%%%%%%%%%

\subsection{Parameterizations of the 7-sphere}

To be explicit, we parametrize the $S^7$ by the coordinates \cite{plefka}
\bal
		z_1
		&=
		\cos\alpha\,\cos\frac{\theta_1}{2}\,e^{i\xi_1}\,,\quad&
		z_2
		&=
		\cos\alpha\,\sin\frac{\theta_1}{2}\,e^{i\xi_2}\,,\\
		z_3
		&=
		\sin\alpha\,\cos\frac{\theta_2}{2}\,e^{i\xi_3}\,,\quad &
		z_4
		&=
		\sin\alpha\,\sin\frac{\theta_2}{2}\,e^{i\xi_4}\,.
\eal
This choice induces the following metric
\bal
	ds^2
	&=
	\frac14
	\left[
		4d\alpha^2
		+
		\cos^2\alpha
		\left(
			d\theta_1^2
			+4\cos^2\frac{\theta_1}{2}\, d\xi_1^2
			+4 \sin^2\frac{\theta_1}{2}\,d\xi_2^2
		\right)
	\right.
	\\&\hskip2cm \left.
		+
		\sin^2\alpha
		\left(
			d\theta_2^2
			+4\cos^2\frac{\theta_2}{2}\,d\xi_3^2
			+4\sin^2\frac{\theta_2}{2}\, d\xi_4^2
		\right)
	\right]
	\,,
\eal
with appropriate ranges for the coordinates to cover the entire sphere once, namely $0\le \alpha\le \pi/2$, $0\le \theta_{1,2}\le \pi$ and $0\le \xi_i\le 2\pi$.

%In the following, it will be useful to consider the 7-sphere in two different coordinate 
%systems, one that makes the action of 
%and one that is more suited for the supersymmetry analysis of the M2-brane embedding. 
%In both cases, we start by defining

To make the action of $\bZ_p$ and $\bZ_q$ more transparent, we 
can relabel the $\xi_i$ as
\beq
\xi_1=\frac{\phi_1}{2}+x\,,\qquad
\xi_2=-\frac{\phi_1}{2}+x\,,\qquad
\xi_3=\frac{\phi_2}{2}+v\,,\qquad
\xi_4=-\frac{\phi_2}{2}+v\,,
\label{sub1}
\eeq
which results in
\be
ds^2_{S^7/(\bZ_p\oplus \bZ_q)}=\frac{1}{4}\left[
4d\alpha^2+\cos^2\alpha\, ds^2_{S^3/\mathbb{Z}_p}+
\sin^2\alpha\, ds^2_{S^3/\mathbb{Z}_q}\right]\,,
\label{metric-spherek1}
\ee
with
\bal
\label{S3-orbifolds}
ds^2_{S^3/\mathbb{Z}_p}&=
d\theta_1^2+d\phi_1^2+4dx^2+4\cos\theta_1\, dx\, d\phi_1\,,\\
ds^2_{S^3/\mathbb{Z}_q}&=
d\theta_2^2+d\phi_2^2+4dv^2+4\cos\theta_2\, dv\, d\phi_2\,.
\eal
We are implicitly assuming that $k=1$, so that $\mathbb{Z}_k$ is trivial. 
The action of $\mathbb{Z}_p$ and $\bZ_q$ then
identify~\cite{imamura,assel2}
\beq
	(z_1,z_2,z_3,z_4)
	\sim
	(\omega^m_p z_1,\omega_p^m z_2,\omega_q^n z_3,\omega_q^n z_4)\,,
	\qquad
	m,n\in\bZ\,,
\eeq
and therefore impose the following periodicities on the angular coordinate $x$ and $v$
\be
x\sim x+\frac{2\pi m}{p}\,,\qquad
v\sim v+\frac{2\pi n}{q}\,.
\label{orb_pq}
\ee
With these coordinates the action of $SU(2)_{A}$ and $SU(2)_B$ is manifest 
on the two $S^3$ orbifolds in \eqn{S3-orbifolds}.

The full metric of the space (before the $\bZ_k$ orbifold) is given by \cite{assel2}
\be
ds^2=R^2\left(ds^2_{AdS_4}+4 ds^2_{S^7/(\bZ_p\oplus \bZ_q)}\right)\,.
\label{metric-correct}
\ee
The Killing spinor for this 11-dimensional geometry is derived in Appendix~\ref{appKS} and is given by
\be
\eta=\Psi\, \Xi\,\eta_0\,,
\label{etafinal}
\ee
with
\bal
\Psi&=
e^{-\tfrac{\alpha-\pi/2}{2} \Gamma_4\Gamma_\star}\,
e^{\tfrac{\theta_1-\pi/2}{4}(\Gamma_{45}+\Gamma_{67})}\,
e^{\tfrac{\theta_2-\pi/2}{4}(\Gamma_{9\#}-\Gamma_{8}\Gamma_\star)}\,
e^{\tfrac{x}{2}(\Gamma_{46}+\Gamma_{57})}
\\
& \hskip 5cm
{}\times 
e^{\tfrac{v}{2}(\Gamma_{8\#}-\Gamma_{9}\Gamma_\star)}\,
e^{-\tfrac{\phi_1}{4}(\Gamma_{47}+\Gamma_{56})}\,
e^{-\tfrac{\phi_2}{4}(\Gamma_{89}-\Gamma_\#\Gamma_\star)}\,
\,,\\
\Xi&=e^{\tfrac{u}{2}\Gamma_2\Gamma_\star}
e^{\tfrac{\rho}{2}\Gamma_1\Gamma_\star}
e^{\tfrac{t}{2}\Gamma_0\Gamma_\star}
e^{\tfrac{\phi}{2}\Gamma_{23}}\,,
\label{etafinal1}
\eal
where $\eta_0$ is a 32-component constant spinor and $\Gamma_\star\equiv \Gamma_{0123}$. 
The gamma matrices live in the tangent space and the indices $0,1,\ldots,9,\#$ are flat indices.

A second parameterization of the angles is useful to study the general case of 
$k\ge 2$ and is obtained by setting
\bea
&\xi_1=\frac{1}{4}\left(-2\phi_1+\chi+\zeta\right),\qquad
\xi_2=\frac{1}{4}\left(2\phi_1+\chi+\zeta\right),\cr
&\xi_3=\frac{1}{4}\left(-2\phi_2-\chi+\zeta\right),\qquad
\xi_4=\frac{1}{4}\left(2\phi_2-\chi+\zeta\right),
\label{sub2}
\eea
in the metric above. 
This results in \cite{plefka}
\bal
ds^2_{M_{p,q,k}}
=&\,
\frac14
\Big[
4d\alpha^2+
\sin^2\alpha\left(d\theta_1^2+\sin^2\theta_1\,d\phi_1^2\right)+\sin^2\alpha\left(d\theta_2^2
+\cos^2\theta_2\,d\phi_2^2\right)
\\ &\hskip 1cm\left.
+\cos^2\alpha\,\sin^2\alpha
\left(d\chi+\cos\theta_1\,d\phi_1-\cos\theta_2\,d\phi_2\right)^2+ \frac{1}{4}\left(d\zeta+A\right)^2
\right]\,,
\label{metricspherezeta}
\eal
where
\beq
A=\cos2\alpha\,d\chi
+2\cos^2\alpha\,\cos\theta_1\,d\phi_1
+2\sin^2\alpha\,\cos\theta_2\,d\phi_2\,.
\eeq
In these coordinates, the action of the $\bZ_k$ orbifold is along the $\zeta$ direction, \emph{i.e.}
\beq
\zeta\sim\zeta+\frac{2\pi m}{k}\,,\qquad m\in\mathbb{Z}\,.
\eeq

The Killing spinor for these coordinates was obtained in \cite{plefka} and is as in (\ref{etafinal}), with $\Psi$ replaced by
\be
\tilde \Psi=e^{\frac{\alpha}{2}(\Gamma_4\Gamma_\star-\Gamma_{7\#})}
e^{\frac{\theta_1}{4}(\Gamma_5\Gamma_\star-\Gamma_{8\#})}
e^{\frac{\theta_2}{4}(\Gamma_{79}+\Gamma_{46})}
e^{-\frac{\xi_1}{2}\Gamma_\# \gamma_\star}
e^{-\frac{\xi_2}{2}\Gamma_{58}}
e^{-\frac{\xi_3}{2}\Gamma_{47}}
e^{-\frac{\xi_4}{2}\Gamma_{69}}
\,,
\label{etafinal2}
\ee
where the $\xi_i$ are given by (\ref{sub2}).

%%%%%%%%%%%%%

\subsection{Supersymmetry analysis}

The task at this point is twofold. First, we want to check that the three orbifolds of 
$M_{p,q,r}$ preserve half of the supersymmetries of the vacuum, namely 16 supercharges. 
Secondly, we want to find an M2-brane embedding that is 1/2 BPS, 
breaking these 16 supersymmetries down to 8.

To address the first point, we consider the first parameterization, equations 
(\ref{metric-spherek1}) and (\ref{etafinal1}) above, and use the procedure 
employed in \cite{plefka} for the ABJM case. This begins by choosing a 
basis for the gamma matrices such that
\beq
\Gamma_{0123456789\#}=1\,,
\eeq
and by writing the constant spinor $\eta_0$ in a basis which block-diagonalizes as follows 
\beq
	\Gamma_{46}\eta_0
	=
	s_1 \eta_0\,,
	\quad
	\Gamma_{57}\eta_0
	=
	s_2 \eta_0\,,
	\quad
	\Gamma_{8\#}\eta_0
	=
	s_3 \eta_0\,,
	\quad
	\Gamma_{9}\Gamma_\star \eta_0
	=
	s_4 \eta_0\,.
\eeq
All $s_i$ eigenvalues take values $\pm1$, see \cite{plefka}. A translation in $x$ corresponds to an action of the $\bZ_p$ orbifold and rescales the Killing spinor by 
\beq
	\Psi\,\Xi\,\eta_0
	\rightarrow
	e^{i(s_1+s_2)\delta/2}\Psi\,\Xi\,\eta_0\,,
\eeq
as can be readily seen from (\ref{etafinal1}). In order for this to be a symmetry of the Killing spinor (for a generic translation parameter $\delta$), we are restricted to
\beq
	(s_1,s_2,s_3,s_4)
	\in
	\left\{
		(+,-,\,\cdot\, ,\,\cdot\,)\,,
		(-,+,\,\cdot\,,\,\cdot\,)
	\right\}\,,
\eeq
where $\pm$ indicates $\pm1$ and $\cdot$ indicates an unrestricted value. The action of the $\bZ_q$ orbifold ({\it i.e.}, a translation in $v$) similarly gives
\beq
	(s_1,s_2,s_3,s_4)
	\in
	\left\{
		(\,\cdot\,,\,\cdot\,,+,+)\,,
		(\,\cdot\,,\,\cdot\,,-,-)
	\right\}\,.
\eeq
Now we notice that $\zeta\propto x+v$, so that the action of the $\bZ_k$ orbifold rescales the Killing spinor
as\beq
	\Psi\,\Xi\,\eta_0
	\rightarrow
	e^{i(s_1+s_2+s_3-s_4)\delta/2}\Psi\,\Xi\,\eta_0\,.
\eeq
This restricts to the following choices
\bal
	(s_1,s_2,s_3,s_4)
	\in&\,
	\left\{
		(+,-,-,-)\,,
		(-,+,-,-)\,,
		(-,-,+,-)\,,
	\right.\\
	&\quad\left.
		(-,+,+,+)\,,
		(+,-,+,+)\,,
		(+,+,-,+)
	\right\}\,.
\eal
Note that each set of $s_i$ eigenvalues represents four supercharges. Thus, 
the $\bZ_k$ orbifold by itself restricts us to 24 supercharges, \emph{i.e.}, the 
ABJM theory, as expected. Taking the intersection of the three sets of allowed 
eigenvalues, we are left with
\bal
	(s_1,s_2,s_3,s_4)\in&\,
	\left\{	(+,-,-,-)\,,(-,+,-,-)\,,(-,+,+,+)\,,(+,-,+,+)
	\right\}\,,
	\label{16options}
\eal
namely 16 supercharges. 

%\subsection{M2-brane embedding}

The supersymmetry analysis of the M2-brane embedding is most easily carried out using the second parametrization, equations (\ref{metricspherezeta}) and (\ref{etafinal2}) above. As anticipated earlier, the M2-brane is taken to be oriented along the $AdS_2$ directions in $AdS_4$ (while sitting at $u=0$) and to wrap the $S^1$ parametrized by $\zeta$. The world-volume coordinates are then given by $(t,\rho,\zeta)$, with the remaining coordinates on the sphere assuming constant values. The induced metric is thus simply given by
\be
\frac{ds^2_\mt{ind}}{R^2}=d\rho^2-\cosh^2\rho\, dt^2+\frac{1}{4}d\zeta^2\,,
\ee
where $R^6=2^5\pi^2 N k p q$, as seen above \cite{imamura}. 
The amount of supersymmetry preserved by this embedding is obtained by considering the projection equation \cite{Skenderis:2002vf}
\be
\Gamma \eta=\eta\,,
\ee
for the Killing spinor (\ref{etafinal2}), with the projector being given by
\be
\Gamma=\frac{1}{\sqrt{-\det g}} \partial_t x^\mu\partial_\rho x^\nu \partial_\zeta x^\sigma \gamma_\mu\gamma_\nu\gamma_\sigma=
\Gamma_{01\#}\,.
\ee
This is formally identical to the ABJM case considered in \cite{plefka}, so that we can quote the 
results from that case. Setting therefore $\alpha=0$ and $\theta_1=0$ the projection equation for 
$\epsilon$ reduces to an equation for $\epsilon_0$ and the brane breaks half of the 
supercharges in (\ref{16options}). 
Selecting $\alpha=\pi/2$ and $\theta_2=0$ also gives $1/2$ BPS embeddings, which are 
the loops invariant under $SU(2)_A$.

Unlike the ABJM case, in the present case there are also the $\bZ_p$ and $\bZ_q$ orbifolds. 
A crucial fact is that $\alpha=0$, where our M2-brane is located, is a fixed point of the 
$\bZ_q$ orbifold. In fact, we expect $q$ different states at the singular point and find 
holographic duals for $q$ different fundamental Wilson loops. 
This mirrors (part of) the 
degeneracy of $1/2$ BPS loop operators we have uncovered in the gauge theory side. 
We expect there to be a single fundamental Wilson loop for every one of the $q$ 
twisted hypermultiplets, represented by a dashed line in the graphs in Section~\ref{sec:sl}. 
It would be advantageous to use the type IIB description 
\cite{dhoker1,dhoker2,assel1,assel2}, where this singularity 
is resolved, and which allows also to study linear quivers. See more in the discussion below.

The computation of the renormalized on-shell action for the M2-branes also follows 
 from the ABJM theory case. 
The brane action (after renormalization) will be proportional to $R^3 \int d\zeta$. Using the explicit expression of the radius written above and that the range of $\zeta$ goes up to $1/kp$, the expectation value 
would then seem to be given by
\be
\vev{W}\simeq e^{\pi\sqrt{\frac{2 N q}{kp}}}\,.
\ee
It is actually unlikely that this expression be valid for all the $q$ different Wilson loops, as 
the answer should depend on the details of the resolution of the orbifold singularity (and fluxes, 
and fractional branes). This expression does in fact hold at least in the simplest
case where the Wilson loops are all equivalent, so theories with $p=q$ and alternating 
$\pm k$ CS levels. Those in fact are orbifolds of the ABJM theory.

%%%%%%%%%%%%%%%%%

\section{Discussion}
\label{sec:disc}

The structure of $\cN=4$ Chern-Simons-matter theories in three dimensions is very rich 
and the story of the $1/2$ BPS Wilson loops in those theories is even richer and 
rather complicated.

We expect that a theory with $p$ untwisted hypermultiplets and $q$ twisted hypermultiplets possesses
$q$ independent $1/2$ BPS Wilson loops (not accounting for possible representations)
preserving our supercharges \eqn{ta}. There are $p$ other Wilson loops preserving the
supercharges with the underlined and overlined indices interchanged. They can be easily written 
explicitly, and one can also study them by looking at a quiver with the untwisted and
twisted hypermultipets exchanged.

In the M-theory picture we found that the Wilson loops are indeed dual to M2-branes
wraping a circle in
$\left(S^7/(\bZ_p\oplus \bZ_q)\right)/\bZ_k$ which is at a fixed point of the $\bZ_q$
orbifold and of circumference proportional to $1/kp$. We expect that there are
in fact $q$ different degenerate states at the singular point, representing the holographic
duals of the $q$ Wilson loops in fundamental representations. To resolve the singularity one
can study the type IIB duals of these theories \cite{dhoker1,dhoker2,assel1,assel2}. 
These also allow to study linear
quivers, while M-theory is only a good description for the circular quivers. It is also easier to
specify the different ranks of the gauge groups and the location of the $k=0$ nodes in the
IIB language. The holographic duals of the Wilson loops will be fundamental strings at $q$
specific points on one of the boundaries of the strip/annulus which appears in the metric.
This mirrors the brane construction of these theories \cite{assel1,assel2} and the possible ways to
add fundamental strings to them \cite{assel_gomis}.%
\footnote{We thank B. Assel and J. Estes for explaining some of these points to us.}

On the field theory side the story is more convoluted. In the simple case of alternating levels
we found $2q$ possible superconnections. A pair for each pair of vector multiplets connected by a
hypermultiplet, coupling to the fermion $\psi_{\ul1}^+$ or to $\psi_{\ul2}^-$. 
Each of those superconnections
is invariant under supersymmetry up to a total derivative, that cancels when we consider the
entire Wilson loop. We expect the degeneracy between those Wilson loops to be lifted by
quantum corrections.

This is a phenomenon that has so far not afflicted the study of BPS Wilson loops, but is in fact
common in other settings. The Wilson loop is a composite operator made of the fields at 
arbitrary points along the curve and whose interactions could lead to quantum effects. The 
variation performed here is classical and does not take into account 
possible interactions between the fields. If there is a unique operator carrying some set 
of quantum numbers (in the case of the Wilson loop it is the contour and representation), then 
the classical calculation should suffice. In our case we found a pair of operators with the 
same quantum numbers that are classically BPS. As a simple analog, consider scalar operators 
in $\cN=4$ SYM (in $3d$ or $4d$). We can take two complex scalar fields $Z$ and $X$ in 
the vector multiplet. $\Tr(Z^J)$ and $\Tr(X^J)$ are both 1/2 BPS operators, but $\Tr(Z^2X^2)$ 
is not. The first two are the only operators of classical dimension $J$ carrying $J$ units of 
charge, so they must be BPS. 
The fields $Z$ and $X$ are each annihilated by half of the supercharges, but there is another 
operator $\Tr(ZXZX)$ which has the same classical dimension and charge as $\Tr(Z^2X^2)$, 
so the classical analysis is not enough and indeed only one linear combination of these is 
BPS once one-loop effects are included.

We expect the same to be true in the case of the Wilson loops and only one combination
(presumably the sum of the loop made of $\cL^{\psi_1}$ and the loop made of $\cL^{\psi_2}$)
to be BPS. This story should hold also for the case of the circle and also for the equivalence
to the $1/4$ BPS loop in Section~\ref{sec:loc}. The matrix model will only calculate the
expectation value of the correct linear combination.

One possible way to check this is to calculate the expectation value of the circular Wilson loops
in perturbation theory. We expect that the Wilson loops with either $\cL^{\psi_1}$ or $\cL^{\psi_2}$
to not be BPS and therefore to suffer from UV divergences. Only the correct linear combination will
be finite and equal to the result of the matrix model calculation.%
\footnote{The Wilson loop found in \cite{wu_zhang} is a linear combination of loops on {\em all}
the nodes of the quiver, and all of them are made of the appropriate $\cL^{\psi_i}$, so we
don't expect it to be BPS.}
This calculation should follow along the lines of the perturbative calculations of the Wilson 
loops in ABJM theory \cite{BGLS1,BGLS2,GMPS}.

Further complications arise when one considers quivers with nodes with vanishing CS levels,
as studied in Section~\ref{sec:k0-k}. In the case studied there, we found five possible
superconnections. Their variation does not vanish and is not a total derivative, but we argued
that the remainder cancels in every expectation value. Four of the connections involve a 
pair of nodes and the last one has a richer structure involving all three nodes (one of them 
doubled). This last one is a superconnection in $U(N_1+N_3|2N_2)$. Indeed it would seem 
natural to also combine the other connections into four possible pairings of block-diagonal 
$U(N_1|N_2)\oplus U(N_2|N_3)\subset U(N_1+N_3|2N_2)$ superconnections.
The resulting Wilson loop are classified by representations of $U(N_1+N_3|2N_2)$ 
and we again expect only one linear combinations to be
BPS. It is not clear actually why a Wilson loop involving only a pair of nodes would 
not be BPS. We leave this question for the future.

It may also be interesting to look for $1/2$ BPS loops in the non-linear sigma 
model of \cite{koh} 
which arise from integrating out all the fields of the $k=0$ nodes.

We have shown that the $1/2$ BPS Wilson loops are cohomologically equivalent to certain 
$1/4$ BPS Wilson loops which can be reduced using localization to simple observables in 
the Chern-Simons-matter matrix models \cite{kapustin}. We leave the evaluation of these 
matrix integrals to future work.

In fact, one way to address the open questions is to look at the structure of the Wilson loops 
in the matrix model. In ABJM theory one can calculate the expectation value of either the 
$1/6$ BPS Wilson loop or the $1/2$ BPS one \cite{putrov,marino,klemm,moriyama}. 
But the calculation of the latter is significantly easier. One can try to evaluate possible 
Wilson loops in the matrix models for arbitrary $\cN=4$ CS-matter theories. For theories 
with equal ranks there is an elegant Fermi-gas formulation of the partition functions 
of those theories \cite{MP2}, which is given to all orders in the $1/N$ expansion 
by an Airy function dependent on two parameters (one of which was found explicitly for 
all $\cN=4$ circular quivers in \cite{Moriyama-Nosaka}). Our expectation is that 
in the alternating CS level case, the simple Wilson loops will be in representations 
of $U(N_I|N_{I+1})$, with a vanishing level $U(N_I+N_{I+2}|2N_{I+1})$, with 
two vanishing levels $U(N_I+2N_{I+2}|2N_{I+1}+N_{I+3})$, etc. Note though, that 
in the case considered in Section~\ref{sec:-k0k}, the $1/2$ BPS loops are simply 
in representations of $U(N_{I+1})$. In fact, it is not clear that the matrix model will 
see a distinction between the $q$ loops we constructed in this paper and the 
$p$ other $1/2$ BPS loops preserving the set of supersymmetries 
invariant under $SU(2)_A$. In fact, we would expect both type of operators to 
give simple observables in the matrix model.

Another approach would be to try to derive the form of the Wilson loops by Higgsing 
part of the gauge groups and finding the resulting operators as was done for the
ABJM theory in \cite{Lee-Lee}.

It would be interesting to also study the transformation of these loops under mirror 
symmetry, where we would expect them to become vortex loop operators 
\cite{DGY}. This was studied in other $3d$ theories in \cite{DOP,assel_gomis}. Note 
also that in the case of the ABJM theory there were $1/3$ BPS vortices preserving 
8 supercharges \cite{DGY}. An analog Wilson loop was never found. Possibly 
a linear combination of two $1/2$ BPS Wilson loops with different couplings 
(the analogs of $\psi_{\ul1}$ and $\psi_{\ul2}$) should be considered the appropriate 
dual.

%%%%%%%%%%%%%%

\subsection*{Acknowledgements}

We are grateful to Benjamin Assel, John Estes, Sanefumi Moriyama and Kostya Zarembo 
for useful discussions.
N.D. would like to thank the hospitality of ICTP-SAIFR, S\~ao Paolo, where this 
project was initiated, and
the IFT, Madrid, during the course of this work.
The research of N.D. is underwritten by an STFC advanced fellowship.
D.T. is supported in part by CNPq and by FAPESP grant 2013/02775-0.

%%%%%%%%%%%%%%%

\appendix

\section{Notation and conventions}
\label{app:not}

We consider the loop in $\bR^{3}$. 
Spinor indices are raised and lowered as
\beq
\psi^a=\epsilon^{ab}\psi_b\,,
\qquad
\psi_a=\epsilon_{ab}\psi^b\,,
\qquad
\epsilon^{+-}=-\epsilon_{+-}=1,
\eeq
and we employ the gamma matrix basis
\beq
(\gamma^\mu)^a{}_b=\{
\sigma^3,\,\sigma^1,\,-\sigma^2\},
\eeq
satisfying $\gamma^\mu\gamma^\nu=\eta^{\mu\nu}+i\epsilon^{\mu\nu\rho}\gamma_\rho$ 
(with $\epsilon^{123}=1$).

As in \cite{im_aux}, we use the epsilon symbol to raise and lower the indices of 
the supersymmetry parameters via 
$\xi^{\bar A\ul B}=\epsilon^{\bar A\bar C}\epsilon^{\ul B\ul D}\xi_{\bar C\ul D}$, with 
$\epsilon^{12}=\epsilon_{12}=1$ for both overlined and underlined indices.
In the main text, we shall always write the supersymmetry parameters with lowered indices.

In this gamma matrix basis, the supersymmetry transformations are given by \cite{im_aux}
\begin{subequations}
\beq
\delta q_{(I)}^{\bar A}
=i\sqrt{2}(\xi^{\bar A\ul B})^a (\psi_{(I)\ul B})_a\,,
\qquad	
\delta \bar q_{(I)\bar A}=
i\sqrt{2}(\xi_{\bar A\ul B})^a (\bar\psi_{(I)}^{\ul B})_a\,,
\eeq
\bal
\delta(\psi_{(I)\ul A})^a=&\,
\sqrt{2}(\slashed D)^a{}_b q^{\bar B}_{(I)}(\xi_{\bar B\ul A})^b
-\frac{\sqrt{2}s_{I}}{k}(\xi_{\bar B\ul A})^a
`\left[\nu_{(I)} q^{\bar B}_{(I)}
-q^{\bar B}_{(I)} \tilde\nu_{(I)}\right]\\
&+\left(\sqrt{2}(\xi_{\bar C\ul B})^a (\varphi_{(I)})^{\ul B}{}_{\ul A} q^{\bar 
C}_{(I)}\right)_{k_{I}=0}
-\left(\frac{2\sqrt{2}s_{I}}{k}(\xi_{\bar C\ul B})^a (\tilde\mu_{(I-1)})^{\ul B}
{}_{\ul A} q^{\bar C}_{(I)}\right)_{k_{I}\neq0}\\
&-\left(\sqrt{2}(\xi_{\bar C\ul B})^a q^{\bar C}_{(I)} (\varphi_{(I+1)})^{\ul B}
{}_{\ul A}\right)_{k_{I+1}=0}\\
&+\left(\frac{2\sqrt{2}s_I}{k}(\xi_{\bar C\ul B})^a q^{\bar C}_{(I)} (\mu_{(I
+1)})^{\ul B}{}_{\ul A}\right)_{k_{I+1}\neq0},
\eal
\bal
\delta(\bar\psi_{(I)}^{\ul A})_a
=&\,\sqrt{2}(\slashed D)_{ab}\bar q^{\bar B}_{(I)}(\xi^{\bar B\ul A})^b
+\frac{\sqrt{2}s_{I}}{k}(\xi^{\bar B\ul A})_a\left[
\tilde\nu_{(I)} \bar q_{(I)\bar B}
-\bar q_{(I)\bar B} \nu_{(I)}\right]\\
&+\left(\sqrt{2}(\xi^{\bar C\ul B})_a \bar q_{(I)\bar C} (\varphi_{(I)})^{\ul A}
{}_{\ul B}\right)_{k_{I}=0}
-\left(\frac{2\sqrt{2}s_{I}}{k}(\xi^{\bar C\ul B})_a \bar q_{(I)\bar C} (\tilde
\mu_{(I-1)})^{\ul A}{}_{\ul B}\right)_{k_{I}\neq0}\\
&-\left(\sqrt{2}(\xi^{\bar C\ul B})_a (\varphi_{(I+1)})^{\ul A}{}_{\ul B} \bar 
q_{(I)\bar C}\right)_{k_{I+1}=0}\\
&+\left(\frac{2\sqrt{2}s_I}{k}(\xi^{\bar C\ul B})_a (\mu_{(I+1)})^{\ul A}{}_{\ul 
B} \bar q_{(I)\bar C}\right)_{k_{I+1}\neq0},
\eal
\bal
\delta A_{(I)\mu}
=&\,-\left(\frac{s_I}{k}(\xi_{\bar A\ul B})^a (\gamma_\mu)_{ab}
\left[(j_{(I)}^{\bar A\ul B})^b
-(\tilde j_{(I-1)}^{\ul B\bar A})^b\right]\right)_{k_I\neq0}\\
&-\left((\xi_{\bar A\ul B})^a (\gamma_\mu)_{ab}
\left((\lambda_{(I)}^{\bar A\ul B})^b
+\frac{s_I}{2k}\left[
(j_{(I)}^{\bar A\ul B})^b+(\tilde j_{(I-1)}^{\bar A\ul B})_b\right]\right)
\right)_{k_I=0},
\eal
\beq
\delta(\varphi_{(I)})^{\ul A}{}_{\ul B}
=2i(\xi_{\bar C\ul B})^a (\lambda_{(I)}^{\bar C\ul A})_a
-\delta^{\ul A}{}_{\ul B} i(\xi_{\bar C\ul D})^a (\lambda_{(I)}^{\bar C\ul D})_a\,,
\eeq
\end{subequations}
where for the hypermultiplet fields the label $(I)$ indicates the link between 
the $I$\textsuperscript{th} and $(I+1)$\textsuperscript{st} node and $s_I=1$ for a hypermultiplet.
The bracketed terms vanish unless the condition in the subscript is satisfied.
To obtain the transformations for the twisted multiplets, we replace the 
overlined $SU(2)_A$ indices for underlined $SU(2)_B$ indices and choose $s_I=-1$.

The superconformal transformations of the fields are found by replacing $\xi^a
\rightarrow x^\mu(\gamma_\mu)^a{}_b\eta^b$, with the exception of the fermions.
The variation of the fermions with respect to the superconformal supercharges are 
given by
\bal
\delta(\psi_{(I)\ul A})^a
=&\,\sqrt{2}(\slashed D)^a{}_b q^{\bar B}_{(I)}x^\mu(\gamma_\mu)^b{}_c
(\eta_{\bar B\ul A})^c
+\sqrt{2}q^{\bar B}_{(I)} \eta_{\bar B\ul A}^a\\
&-\frac{\sqrt{2}s_{I}}{k}x^\mu(\gamma_\mu)^a{}_b(\eta_{\bar B\ul A})^b
\left[(\nu_{(I)})^{\bar C}{}_{\bar C} q^{\bar B}_{(I)}
-q^{\bar B}_{(I)} (\tilde\nu_{(I)})^{\bar C}{}_{\bar C}\right]\\
&+\left(\sqrt{2}x^\mu(\gamma_\mu)^a{}_b(\eta_{\bar C\ul B})^b 
(\varphi_{(I)})^{\ul B}{}_{\ul A} q^{\bar C}_{(I)}\right)_{k_{I}=0}\\
&-\left(\frac{2\sqrt{2}s_{I}}{k}x^\mu(\gamma_\mu)^a{}_b(\eta_{\bar C\ul B})^b 
(\tilde\mu_{(I-1)})^{\ul B}{}_{\ul A} q^{\bar C}_{(I)}\right)_{k_{I}\neq0}\\
&-\left(\sqrt{2}x^\mu(\gamma_\mu)^a{}_b(\eta_{\bar C\ul B})^b q^{\bar C}_{(I)} 
(\varphi_{(I+1)})^{\ul B}{}_{\ul A}\right)_{k_{I+1}=0}\\
&+\left(\frac{2\sqrt{2}s_I}{k}x^\mu(\gamma_\mu)^a{}_b(\eta_{\bar C\ul B})^b 
q^{\bar C}_{(I)} (\mu_{(I+1)})^{\ul B}{}_{\ul A}\right)_{k_{I+1}\neq0},
\eal
\bal
\delta(\bar\psi_{(I)}^{\ul A})_a
=&\,\sqrt{2}(\slashed D)_{ab}\bar q^{\bar B}_{(I)}x^\mu(\gamma_\mu)^b{}_c 
\eta^{\bar B\ul A})^c
+\sqrt{2}\bar q_{(I)\bar B}\eta^{\bar B\ul A}_a\\
&+\frac{\sqrt{2}s_{I}}{k}x^\mu(\gamma_\mu)^a{}_b \eta^{\bar B\ul A})^b
\left[(\tilde\nu_{(I)})^{\bar C}{}_{\bar C} \bar q_{(I)\bar B}
-\bar q_{(I)\bar B} (\nu_{(I)})^{\bar C}{}_{\bar C}\right]\\
&+\left(\sqrt{2}x^\mu(\gamma_\mu)^a{}_b \eta^{\bar C\ul B})^b \bar q_{(I)\bar C} 
(\varphi_{(I)})^{\ul A}{}_{\ul B}\right)_{k_{I}=0}\\
&-\left(\frac{2\sqrt{2}s_{I}}{k}x^\mu(\gamma_\mu)^a{}_b \eta^{\bar C\ul B})^b 
\bar q_{(I)\bar C} (\tilde\mu_{(I-1)})^{\ul A}{}_{\ul B}\right)_{k_{I}\neq0}\\
&-\left(\sqrt{2}x^\mu(\gamma_\mu)^a{}_b \eta^{\bar C\ul B})^b 
(\varphi_{(I+1)})^{\ul A}{}_{\ul B} \bar q_{(I)\bar C}\right)_{k_{I+1}=0}\\
&+\left(\frac{2\sqrt{2}s_I}{k}x^\mu(\gamma_\mu)^a{}_b \eta^{\bar C\ul B})^b 
(\mu_{(I+1)})^{\ul A}{}_{\ul B} \bar q_{(I)\bar C}\right)_{k_{I+1}\neq0}.
\eal

%%%%%%%%%%%%%%%%%%%

\section{Derivation of the Killing spinor (\ref{etafinal1})}
\label{appKS}

In this appendix we work out the explicit expression for the Killing spinor associated to the first parameterization of the space employed in the main text, namely equations (\ref{metric-spherek1}) and (\ref{metric-correct}).

The AdS part is standard. For example, for a time-like line the vielbeins can be chosen as follows
\be
e^0= R \cosh u\, \cosh \rho\, dt\,,\quad
e^1= R \cosh u\, d\rho\,,\quad
e^2= R \, du\,,\quad
e^3= R \sinh u\, d\phi\,,
\ee
and the relevant non-vanishing components of the spin connection are given by
\be
\omega^{01}=\sinh\rho\, dt\,, \quad \omega^{02}=\sinh u\, \cosh\rho\, dt\,,\quad
\omega^{12}=\sinh u\, d\rho\,,\quad \omega^{32}=\cosh u\, d\phi\,,
\ee
The indices $0,1,\ldots,9,\#$ are on the tangent space and in the following we denote them by the letters from the beginning of the Latin alphabet $a,b,\ldots$. We can take the background 4-form field-strength to be proportional to the volume form of $AdS_4$
\be
H_{(4)}
=\kappa \,e^0\wedge e^1\wedge e^2\wedge e^3\
=\kappa R^4 \cosh^2u\, \cosh\rho\, \sinh u\, dt\wedge d\rho\wedge du\wedge d\phi
\,,
\ee
with $\kappa$ a constant to be determined. The Killing spinor equation is given by the variation of the gravitino and reads \cite{Lu:1998nu}
\be
D_\mu \eta+\frac{1}{288}\left(\gamma_{\mu\nu\rho\sigma\tau}H^{\nu\rho\sigma\tau}-8
H_{\mu\nu\rho\sigma}\gamma^{\nu\rho\sigma}\right)\eta=0\,,
\label{KSequation}
\ee
where $D_\mu=\partial_\mu+\tfrac{1}{4}\omega_\mu^{ab}\Gamma_{ab}$ and $\gamma_\mu=e_\mu^a\Gamma_a$ (lower case $\gamma$'s are in the curved space, while upper case $\Gamma$'s are in the tangent space). 
Now we distinguish between the AdS and the sphere directions. Using the expression for the 4-form we arrive at
\bea
&&D_\mu \eta=
-\frac{\kappa}{6}\gamma_\mu\Gamma_\star \eta\,,\qquad \mu \mbox{ along } AdS_4\,,\cr
&&D_\mu \eta=
\frac{\kappa}{12}\gamma_\mu\Gamma_\star \eta\,,\qquad \;\, \mu \mbox{ along } S^7\,,
\eea
with $\Gamma_\star\equiv \Gamma_{0123}$. To find the solution it is useful to use the following identities. If $[X,Y]=2Z$ and $[X,Z]=-2Y$ then
\be
e^{\tfrac{1}{2}\theta X}Y =\left(\cos\theta\, Y+\sin\theta\, Z\right) e^{\tfrac{1}{2}\theta X}\,.
\label{id1}
\ee
If, on the other hand, $[X,Y]=\pm 2Z$ and $[X,Z]=\pm 2Y$ then
\be
e^{\tfrac{1}{2}\theta X}Y =\left(\cosh\theta\, Y\pm \sinh\theta\, Z\right) e^{\tfrac{1}{2}\theta X}\,.
\label{id2}
\ee
The $AdS_4$ part of the Killing spinor is given by
\bea
e^{\tfrac{u}{2}\Gamma_2\Gamma_\star}
e^{\tfrac{\rho}{2}\Gamma_1\Gamma_\star}
e^{\tfrac{t}{2}\Gamma_0\Gamma_\star}
e^{\tfrac{\phi}{2}\Gamma_{23}}\eta_0\,.
\eea 
It can be readily checked that it must be $\kappa=-3/R$. 

On the other hand, the vielbeins for the 7-sphere (\ref{metric-spherek1}) are given by
\bea
&& e^4= 2R \,d\alpha\,,\cr
&& e^5= - R \cos\alpha\, d\theta_1\,,\quad
e^6= -R \cos\alpha \left(2dx+\cos\theta_1\,d\phi_1\right)\,,\quad
e^7= R \cos\alpha \,\sin\theta_1\,d\phi_1\,,\cr
&& e^8= R \sin\alpha\, d\theta_2\,,\quad
e^9= R \sin\alpha \left(2dv+\cos\theta_2\,d\phi_2\right)\,,\quad
e^\#= -R \sin\alpha \,\sin\theta_2\,d\phi_2\,,\cr &&
\eea
and the non-vanishing components of the spin connection are
\bea
&& \omega^{54}=\tfrac{1}{2}\sin\alpha\,d\theta_1\,,\quad
\omega^{64}=\sin\alpha\left(dx+\tfrac{1}{2}\cos\theta_1\,d\phi_1\right)\,,\quad
\omega^{74}=-\tfrac{1}{2}\sin\alpha\, \sin\theta_1\, d\phi_1\,,\cr
&& \omega^{48}=-\tfrac{1}{2}\cos\alpha\, d\theta_2\,,\quad
\omega^{49}=-\cos\alpha\left(dv+\tfrac{1}{2}\cos\theta_2\, d\phi_2\right)\,,\quad
\omega^{4\#}=\tfrac{1}{2}\cos\alpha\,\sin\theta_2\,  d\phi_2\,,\cr
&&\omega^{65}=-\tfrac{1}{2}\sin\theta_1\, d\phi_1\,,\quad
\omega^{75}=dx-\tfrac{1}{2}\cos\theta_1\,d\phi_1\,,\quad
\omega^{76}=\tfrac{1}{2}d\theta_1\,,\cr
&&
\omega^{98}=-\tfrac{1}{2}\sin\theta_2\,d\phi_2\,,\quad
\omega^{\# 8}=dv-\tfrac{1}{2}\cos\theta_2\,d\phi_2\,,\quad
\omega^{\# 9}=\tfrac{1}{2}d\theta_2\,.
\eea
The Killing spinor equations for the sphere components are
\bea
\partial_\alpha\eta&=&
-\frac{1}{2} \Gamma_4\Gamma_\star\eta\,,
\cr
\partial_{\theta_1}\eta&=&
\frac{1}{4}\left(
\sin\alpha\, \Gamma_{45}+\Gamma_{67}+\cos\alpha\,\Gamma_5\Gamma_\star
\right)\eta\,,
\cr
\partial_x\eta&=&\frac{1}{2}\left(
\sin\alpha\, \Gamma_{46}+\Gamma_{57}+\cos\alpha\,\Gamma_6\Gamma_\star
\right)\eta\,,
\cr
\partial_{\phi_1}\eta&=&-
\frac{1}{4}\Big(\sin\alpha\left(
\cos\theta_1\, \Gamma_{64}-\sin\theta_1\, \Gamma_{74}\right)+\sin\theta_1\,\Gamma_{56}+\cos\theta_1\,\Gamma_{57}
\cr
&& \hskip 4cm
-\cos\alpha\left(\cos\theta_1\Gamma_6\Gamma_\star-\sin\theta_1\,\Gamma_7\Gamma_\star
\right)\Big)\eta\,,
\cr
\partial_{\theta_2}\eta&=&
-\frac{1}{4}\left(
\cos\alpha\, \Gamma_{84}+\Gamma_{9\#}+\sin\alpha\,\Gamma_8\Gamma_\star
\right)\eta\,,
\cr
\partial_v\eta&=&-
\frac{1}{2}\left(
\cos\alpha\, \Gamma_{94}+\Gamma_{8\#}+\sin\alpha\,\Gamma_9\Gamma_\star
\right)\eta\,,
\cr
\partial_{\phi_2}\eta&=&-
\frac{1}{4}\Big(-\cos\alpha\left(
\cos\theta_2\, \Gamma_{49}-\sin\theta_2\, \Gamma_{4\#}\right)+\sin\theta_2\,\Gamma_{89}+\cos\theta_2\,\Gamma_{8\#}
\cr
&& \hskip 4cm
+\sin\alpha\left(\cos\theta_2\Gamma_9\Gamma_\star-\sin\theta_2\,\Gamma_\#\Gamma_\star
\right)\Big)\eta\,.
\label{S7equations}
\eea
With a little bit of algebra it can be checked that $\eta=\Psi\,\eta_0$, with $\Psi$ given by (\ref{etafinal1}), is indeed a solution of these equations.

%%%%%%%%%%%%%

\bibliography{N4WLref}
\end{document}